\documentclass[useAMS,usenatbib]{mn2e}
\usepackage{graphicx}
\usepackage{amsmath}
\usepackage[T1]{fontenc}
\usepackage{aecompl}

\title[H-ATLAS/GAMA: The Morphological Transformation
of Galaxies]{H-ATLAS/GAMA: Quantifying the Morphological Evolution of the Galaxy
Population Using Cosmic Calorimetry}
\author[S.A. Eales]{Stephen Eales$^{1}$\thanks{E-mail:
sae@astro.cf.ac.uk}, Andrew Fullard$^{1}$, Matthew Allen$^{1}$,M.W.L. Smith$^1$, 
\newauthor
Ivan Baldry$^2$,
Nathan Bourne$^3$, C.J.R. Clark$^1$, Simon Driver$^4$,Loretta Dunne$^{3,5}$, 
\newauthor
Simon Dye$^6$, Alister W. Graham$^7$, Edo Ibar$^8$, Andrew Hopkins$^9$, 
\newauthor
Rob Ivison$^{3,10}$, Lee S. Kelvin$^{11}$, Steve Maddox$^{3,5}$, Claudia Maraston$^{12}$,
\newauthor
Aaron S.G. Robotham$^{13}$, Dan Smith$^{14}$, Edward N. Taylor$^{15}$, Elisabetta Valiante$^1$, 
\newauthor
Paul van der Werf$^{16}$, Maarten Baes$^{17}$,
Sarah Brough$^9$, David Clements$^{18}$, 
\newauthor
Asantha Cooray$^{19}$, Haley Gomez$^1$, Jon Loveday$^{20}$, Steven Phillipps$^{21}$,
Douglas Scott$^{22}$ 
\newauthor
and Steve Serjeant$^{23}$\\
$^{1}$School of Physics and Astronomy, Cardiff University, The Parade, Cardiff CF24 3AA, UK\\
$^2$ Astrophysics Research Institute, Liverpool John Moores University, IC2,\\ 
Liverpool Science Park, 146 Brownlow Hill, Liverpool L3 3RF, UK\\
$^3$ Institute for Astronomy, The University of Edinburgh, Royal Observatory, Blackford Hill,
Edinburgh, EH9 3HJ, UK\\
$^4$ International Centre for Radio Astronomy Research, 7 Fairway, The University of Western\\
Australia, Crawley, Perth, WA 6009, Australia\\
$^5$ Department of Physics and Astronomy, University of Canterbury, Christchurch, New Zealand\\
$^6$ School of Physics and Astronomy, University of Nottingham, University Park, Nottingham
NG7 2RD, UK\\
$^7$ Centre for Astrophysics and Supercomputing, Swinburne University of Technology, Hawthorn,
Victoria 3122, Australia\\
$^8$ Instituto de F\'isica y Astronom\'ia, Universidad de Valpara\'iso, Avda. Gran Breta\~na 1111, 
Valpara\'iso, Chile\\
$^9$ Australian Astronomical Observatory, PO Box 915, North Ryde, NSW 1670, Australia\\
$^{10}$ European Southern Observatory, Karl-Schwarzschild-Strasse 2, 85748, Garching, Germany\\
$^{11}$ Institut fur Astro- und Teilchenphysik, Universitat Innsbruck, Technikerstrasse 25, A-6020
Innsbruck, Austria\\
$^{12}$ Institute of Cosmology and Gravitation (ICG), University of Portsmouth, Dennis Sciama
Building, Burnaby Road,\\ 
Portsmouth PO1 3FX, UK\\
$^{13}$ Scottish Universities Physics Alliance, School of Physics and Astronomy, University of St.
Andrews, North Haugh,\\
St. Andrews KY16 9SS, UK\\
$^{14}$ Centre for Astrophysics Research, Science and Technology Research Institute, University of
Hertfordshire, Hatfield,\\ 
Herts, AL10 9AB, UK\\
$^{15}$ School of Physics, The University of Melboure, Parkville, VIC 3010, Australia\\
$^{16}$ Leiden Observatory, PO Box 9513, 2300 RA Leiden, the Netherlands\\
$^{17}$ Sterrenkundig Observatorium,Universiteit Gent, Krijgslaan 281 S9, B-9000 Gent, Belgium\\
$^{18}$ Astrophysics Group, Imperial College, Blackett Laboratory, Prince Consort Road, London
SW7 2AZ\\
$^{19}$ Department of Physics and Astronomy, University of California, Irvine, CA 92697, USA\\
$^{20}$ Department of Physics and Astronomy, University of Sussex, Falmer Campus, Brighton BN1 9QH
UK\\
$^{21}$ Astrophysics Group, Department of Physics, University of Bristol, Tyndall Avenue,
Bristol BS8 1TL\\
$^{22}$ Department of Physics and Astronomy, University of British Columbia, 6224 Agricultural
Road, Vancouver,\\
BC V6T 1Z1, Canada\\
$^{23}$ Department of Physical Science, The Open University, Milton Keynes, MK7 6AA, UK
\\ \\ \\ \\ \\ \\ \\ \\ \\ \\ \\ \\ \\ \\ \\ \\ \\ \\ \\ \\ \\ \\ \\ \\ \\ \\
}
\begin{document}

\date{2015 February 10}

\pagerange{\pageref{firstpage}--\pageref{lastpage}} \pubyear{2002}

\maketitle

\label{firstpage}

\begin{abstract}
\newpage
Using results from the Herschel Astrophysical Terrahertz Large-Area
Survey and the Galaxy and Mass Assembly project, we show that, for
galaxy masses above $\simeq10^8\ M_{\odot}$, 51\% of the stellar mass-density
in the local Universe is in early-type galaxies (ETGs: S\'ersic $n >
2.5$) while 89\% of the rate of production of stellar mass-density is
occurring in late-type galaxies (LTGs: S\'ersic $n < 2.5$). From
this
zero-redshift benchmark, we have used a calorimetric
technique to
quantify the importance of the morphological transformation 
of galaxies over the history of the Universe.
The extragalactic background radiation contains 
all the energy generated by nuclear fusion in stars
since the Big Bang. By resolving this background radiation into individual galaxies
using the deepest far-infrared survey with the Herschel Space Observatory and a deep 
near-infrared/optical survey with the Hubble Space Telescope (HST), and using measurements
of the S\'ersic index of these galaxies derived from the HST images, we estimate that
$\simeq$83\% of the stellar mass-density formed over the history of the Universe occurred
in LTGs. 
The difference between this and the fraction of 
the stellar mass-density that is in LTGs today
implies there must have
been a major transformation of LTGs into ETGs
after the formation of most of the stars.

\end{abstract}


\section{Introduction}

Over the last decade much of the research on galaxy evolution has 
started from the paradigm that there are two types of galaxy: star-forming
galaxies and passive or quiescent galaxies in which stars are no
longer forming. In this paradigm star-forming galaxies lie on the Galaxy
Main Sequence (GMS) until some catastrophic process (or at least, a process
brief compared to the
age of the Universe) quenches the star formation, causing the galaxy
to move away from the GMS
and become a `red and dead' galaxy
(e.g. Peng et al. 2010; Speagle et al. 2014).

The rapid evolution of the stellar mass function of passive galaxies 
since $z \sim 1-2$, relative to the much slower evolution
of the stellar mass function of star-forming galaxies 
(Faber et al. 2007; Muzzin et al. 2013; Ilbert et al.
2013), 
implies that the formation of passive galaxies
occurred after the epoch at which the star-formation rate
in the Universe was at its peak (Hopkins and Beacom 2006).
This temporal sequence is circumstantial evidence that
passive galaxies were formed from
the star-forming population. A more
quantitative argument is that
Peng et al. (2010) have shown that the shapes of the two stellar mass functions
are exactly what one would predict from a simple model in which passive galaxies are formed
by the quenching of the star formation in 
the galaxies in the star-forming population.

In an accompanying paper (Eales et al. 2015), we show that when this 
paradigm is considered in the
light of the galaxies detected by the Herschel Space Observatory (Pilbratt et al.
2010) it looks less convincing. Approximately 25\% of the
galaxies detected at low redshift by the largest Herschel extragalactic
survey, the Herschel ATLAS (Eales et al. 2010), are galaxies that would have
been classified as quiescent or passive using optical criteria, but still have
large reservoirs of interstellar material and are still forming stars. They are
red but not dead. The blurring of this dichotomy
when
viewed from a far-infrared perspective is, we argue, evidence
that
some more gradual physical processes (rather than 
catastrophic quenching)
are necessary to explain the properties of the galaxy population (Eales et al. 2015).
At the very least, our results show that there is a major practical problem in measuring star-formation
rates in galaxies and in separating galaxies into the two classes, which may explain the
huge diversity in measurements of the shape and evolution of the GMS found
by Speagle et al. (2014) in their meta-analysis of 25 separate studies of the
GMS.

In this paper we consider a different dichotomy: that between early-type
and late-type galaxies (henceforth ETGs and LTGs).
The terms `early type' and `late type' were coined by Hubble
(Hubble 1926, 1927), although he did not mean the terminology to
indicate that ETGs evolve into LTGs
(Baldry 2008). ETGs are galaxies with a prominent
spheroid or bulge, in particular classes E and S0 in the Hubble
morphological classification, whereas
the structures of LTGs are dominated by a spiral
pattern. We discuss below
how 
to turn this rather vague definition into one that is more
quantitative. 
Although there is
an obvious dichotomy between elliptical galaxies and late-type spirals
when looking at optical images,
the dichotomy between ETGs and LTGs is much less obvious when galaxies are studied more
closely. Integral-field kinematic studies have shown that ETGs
can be split into slow and fast rotators, with the kinematic
properties of the fast rotators being similar to those of LTGs
and the slow rotators being much
rarer than the fast rotators (Emsellem et al. 2011). Careful
quantitative
analysis of the images of ETGs
also often reveals that the galaxy has a disk
even when this is not apparent from simply looking at
the image
(Krajnovic et al. 2013). Observations with the Herschel Space Observatory have
revealed that $\simeq$50\% of ETGs contain dust,
and that there is gradual change of the dust properties of galaxies
as one moves along the Hubble Sequence from LTGs
to ETGs, rather than a sudden jump at the transition
from late types to early types (Smith et al. 2012a).

Although Hubble did not draw any conclusions about the evolutionary
links between ETGs and LTGs, there are many theoretical arguments
for why galaxy evolution is expected to include the morphological
transformation of galaxies. Almost four decades ago, Toomre
(1977) suggested that the merger of two LTGs might scramble
the stars and velocity fields, leading to the formation of
an ETG supported against gravity by the random velocities of the
stars rather than by the rotation of a galactic disk.
On the other hand, if one starts with a spheroid, it seems likely
that the accretion of gas and angular momentum from the intergalactic medium is likely
to lead to the growth of a disk (Combes et al. 2014 and references therein).
A recent idea is that 
a spheroid is formed by the rapid motion of star-forming
clumps towards the centre of a disk (Noguchi
1999; Bournaud et al. 2007; Genzel et al. 2011, 2014;
Dekel and Burkert 2014), 
transforming a LTG into an ETG, with the growth of the
spheroid eventually stopping the formation of stars
in the disk
once the spheroid
has enough mass to stabilize the disk 
against gravitational collapse (Martig et
al. 2009).

There is substantial overlap between ETGs and the class of passive
galaxies and between LTGs and the class of star-forming galaxies.
The overlap between the classes in the low-redshift Universe
is well known (Kennicutt 1998), but over the last decade
a large number of observational programmes with the Hubble Space Telescope
(HST)
have shown that this is also true at high redshift: 
high-redshift passive galaxies tend to have spheroidal
structures and the structures of high-redshift star-forming galaxies 
are generally dominated by disks (Bell et al. 2004a; Wuyts et al. 2011; 
Bell et al. 2012;
Bruce et al. 2012; Buitrago et al. 2013; Szomoru et al. 2013;
Tasca et al. 2014).
These studies show that the evolutionary process that produces passive
galaxies must be linked to the evolutionary process that generates
spheroids (Bell et al. 2012).

These HST studies have two limitations.
The first limitation is a consequence of
the
extragalactic background radiation and of the nature of
the sources which constittue it.
The extragalactic background radiation (Dole et al.
2006; Dwek and Krennrich 2013) is dominated
by radiation of roughly equal strength  in two wavebands, in the optical/near-infrared
and in the far-infrared at $\rm \simeq100-200\ \mu m$.
This equality implies that
$\simeq$50\% of the energy emitted by stars since the Big Bang
has been absorbed by dust and is now in the far-infrared band.
This is an average value and studies of the relative cosmic evolution of
the far-infrated and UV luminoisity density 
(Takeuchi, Buat and Burgarella 2005; Burgarella
et al. 2013) imply that the fraction of the stellar
light absorbed by dust is much higher at $z > 1$ during
the epoch in which most of the stars in the Universe were formed.
An important difference between the 
two peaks is that the sources that constitute
the far-infrared background are rarer and consequently more
luminous. We show in this paper that in a small area
of sky the optical/near-infrared background is composed of
$\simeq$8000 sources but the far-infrared background, which
is of roughly equal strength, is composed of only $\simeq$1500 sources,
implying that the sources of the far-infrared background are
a factor of $\simeq$5 more luminous than the more common sources
composing the optical/near-infrared background.

All of the HST studies above start from optical/near-infrared
samples of galaxies. They all attempt to allow for the
star-formation in each galaxy
that is hidden by dust by using 
either
the optical/near-infrared colours of the galaxy to estimate
the extinction or, in two cases,
the Spitzer 24-$\mu$m observations to estimate the far-infrared
emission.
The probem with the first method is that the luminous galaxies that make
up the far-infrared background are so shrouded by dust that
estimates of their star-formation rates 
from short-wavelength measurements are likely to be inaccurate.
The problem with the second is the significant extrapolation necessary
from the mid-infrared to the far-infrared.

The second limitation of these studies  is that all only reach
conclusions
about the morphologies of the galaxies in restricted samples,
with the usual limits being redshift and stellar mass. As one
example,
Bruce et al. (2012) concluded that the star-forming galaxies in
the redshift range $2 < z < 3$ and with stellar masses $>10^{11}\ M_{\odot}$
have structures that are dominated by disks. 
Although conclusions like this are important, they always leave open
the possibility that most of the stars in the Universe were
formed in a different range of redshift or in galaxies with stellar
masses outside the limits of the sample.
Our overall goal in this paper is to reach a more general conclusion about
the fraction of the stellar mass-density 
that was formed in galaxies of different morphological types over the
entire history of the Universe.

To reach this conclusion, we
have used a calorimetry technique, which is not new (Pagel et al. 1997;
Lilly and Cowie 1987) but is particularly
timely because of one of the
successes of the Herschel Space Observatory. The method
is based on the fact that the
extragalactic background
radiation is the repository of all the radiation emitted as
the result of nuclear fusion in stars.
For the method to be practical,
it is necessary to resolve this background radiation into
individual sources.
Deep surveys with the Hubble Space Telescope
had already resolved the optical/near-infrared background
into individual sources, and now the deepest Herschel survey
has also achieved this goal
in the far infrared
(Elbaz et al. 2011; Magnelli et al. 2013; Leiton et al. 2015).

One of the advantages of the calorimetry technique is that it
does not require us to estimate the star-formation rate
in individual galaxies, which is difficult enough even for galaxies
at zero redshift (Kennicutt 1998; Kennicutt and Evans 2012).
For example, one method of estimating the star-formation rate
in a galaxy is by assuming the far-infrared emission represents
the bolometric luminosity of newly formed stars
(Kennicutt 1998; Kennicutt and Evans 2012). This is probably
a reasonable assumption for the luminous galaxies that
constitute the far-infrared background but
it is clearly wrong in galaxies with low star-formation rates, in which
much of the far-infrared emission is from dust heated by
the old stellar population (Bendo et al. 2015). The calorimetry
method avoids the need to make assumptions like this
by linking the total energy emitted by a galaxy to the
energy produced by nuclear fusion.

Our secondary goal in this paper is to determine how the stellar mass-density
in the Universe today depends on galaxy morphology. By comparing
this distribution with the result from the caloriometry technique
on the morphologies of the galaxies in which most stars were formed, it is
possible to quantify how important the transformation
of galaxy morphology has been in the evolution of the
galaxy population.

We have achieved this secondary goal
using two new wide-field surveys of the
nearby Universe. The Herschel Astrophysical Terrahertz Large Area Survey
(the Herschel-ATLAS or H-ATLAS, Eales et al. 2010) is
the Herschel extragalactic survey covering the largest area of
sky, 550 square degrees, roughly one eightieth of the entire
sky. The survey consists of images at 100, 160, 250, 350 and
500 $\mu$m of five fields, two large fields at the North
and South Celestial Poles (NGP and SGP) and three smaller fields on the
celestial equator. The NGP and the equatorial fields were covered
in the Sloan Digital Sky Survey, but more importantly the equatorial
fields have also been covered by a much deeper spectroscopic survey, the Galaxy
and Mass Assembly (GAMA) redshift survey (Driver et al. 2011).
The GAMA team has produced many different data products for the
galaxies in their survey, and we have made liberal use of these
data products in our investigation.

One of the challenges in an investigation like this is to find
a quantitative measure of galaxy morphology that can be used over
a large range of redshift.
The obvious way to measure the position
of a galaxy on the morphological spectrum from early-type to late-type
is to measure the relative proportions of the stellar mass that is
in the bulge/spheroid and the disk. However, there are two major practical
problems with doing this. The first is that although
for many years it was believed that the spheroids
of galaxies are described by the de Vauclouleur $R^{1/4}$ model (de Vaucouleurs
1948), it
is now realised that galaxy bulges follow a range of intensity profiles
(Graham 2013 and references therein). Second, estimates of the proportion of the galaxy's
stellar mass
that is in the bulge depend strongly on the assumptions made about the
dust extinction within the galaxy (Driver et al. 2008;
Graham and Worley 2008), which requires great care and spatial resolution
to measure accurately.

We have instead used the simpler approach of using the S\'ersic
index as a measure of a galaxy's morphology. The S\'ersic
function is

\smallskip
\begin{align}
I(r) = I_e exp( -b_n[ ({r \over r_e})^{1/n} - 1 ] )
\end{align}
\smallskip

\noindent in which $I_e$ is the intensity at the effective radius, $r_e$,
that contains half of the total light from the model, $n$ is the
S\'ersic index and $b_n$ is a function that depends on $n$
(S\'ersic 1963; Graham and Driver 2005).
This useful function can be used to fit the structures of all galaxy types. 
If the value of the S\'ersic index is one, the S\'ersic profile becomes
an exponential intensity profile, which in the low-redshift
Universe is the intensity profile followed by a disk.
If the value of the S\'ersic index is four, the S\'ersic profile
becomes 
the de Vauclouleur $R^{1/4}$ profile, which is the intensity profile
followed by ellipticals in the low-redshift Universe.
The advantage of adopting this method of measuring the
structure of a galaxy compared to measuring the bulge-to-disk
ratio
is that it is much simpler, not requiring any
assumptions about the
effect of dust extinction or the bulge profile (Graham and Worley
2008).

In Section 2, we describe our investigation of how the
stellar mass-density and
star-formation rate per unit
comoving volume depend on galaxy
morphology in the Universe today.
Section 3 describes the caloriometry method that we have used to
estimate how the mass of stars that has formed over the history
of the Universe depends on the morphology of the galaxy in which
those stars were formed. Section 4 gives the results of the calorimetry
method. The results and their implications are discussed in \S 5 and
the conclusions are given in \S 6.   
We assume the cosmological parameters given from the 
Planck 2013 cosmological analysis (Planck Collaboration XVI 2014): 
a spatially-flat universe
with $\Omega_M = 0.315$ and a Hubble constant of 67.3 $\rm km\ s^{-1}\ Mpc^{-1}$.

\section[]{The Universe today}

\subsection{The relationship between star-formation density and morphology}

In order to carry out a census of star formation, it seems sensible
to start from a far-infrared
survey, since newly formed stars are hidden by dust from optical telescopes.
This procedure might lead to the star-formation rate being underestimated
if there is a population of galaxies in which stars are forming but in which
there is
little dust. However, we show in an accompanying paper that at low redshifts
we do not appear to be missing any star-forming but dust-free galaxies (Eales et
al. 2015).

We have therefore used as our starting point the far-infrared H-ATLAS survey
rather than the optical GAMA survey. We have used the data for the H-ATLAS
equatorial fields (Eales et al. 2010), which are also covered by the GAMA survey
and have a total area of $\simeq$160 deg$^2$. 
The version of the H-ATLAS
data used in this paper is the Phase
1 version 3 internal data release, which was obtained 
using very similar methods to those used for the H-ATLAS
Science Demonstration Phase, rather than the public data
release described by Valiante et al. (2015) and Bourne et al.
(2015). The maps at 250, 350 and 500 $\mu$m were made using
the methods described by Pascale et al. (2011) and the maps
at 100 and 160 $\mu$m were made using the method described in
Ibar et al. (2010). The detection of the sources was performed on
the 250-$\mu$m images using the methods described in Rigby et
al. (2011). Flux measurements at 350 and 500 $\mu$m were 
generally made by measuring the fluxes at the 250-$\mu$m positions
on the 350-$\mu$m and 500-$\mu$m images after convolving the 
latter images with the point spread function of the telescope.
We measured fluxes on the less sensitive 100 and 160-$\mu$m 
images using aperture photometry, with an aperture diameter
chosen to maximise the signal-to-noise for point sources. For
the small percentage of H-ATLAS sources associated with
nearby galaxies, we measured flux densities by carrying out
aperture photometry at all wavelengths with an aperture 
diameter chosen to recover, as nearly as possible, all the flux.

The Phase 1 catalogue consists of the 109,231 sources
detected at $>5\sigma$ at either 250, 350 or 500 $\mu$m (the maps at
100 and 160 $\mu$m are much less sensitive than those at longer
wavelengths). These signal-to-noise limits corresponds to
flux densities at 250, 350 and 500 $\mu$m of 32, 36 and 45 mJy,
respectively. 

As in the Science Demonstration Phase, 
we used a
likelihood-ratio analysis (Smith et al. 2011) 
to match the sources to galaxies in
the Sloan Digital Sky Survey (SDSS, DR7; Abazajian et al.
2009) with $r < 22.4$. With this method, we can
estimate the probability (the reliability - $R$) of a galaxy 
being the true association of a submillimetre source. We treat
galaxies with $R > 0.8$ as being likely counterparts to the
submillimetre sources. There are counterparts with $R > 0.8$
for 29,761 of the 104,657 sources detected at $5\sigma$ at 250 $\mu$m.
Of these counterparts 14,917 have spectroscopic 
redshifts, mostly from the GAMA redshift survey but some
from the SDSS. For the counterparts, there is matched-
aperture photometry in u, g, r, i and z from the SDSS; in Y,
J, H and K from the UKIRT Large Area Survey, which is
part of the UKIRT Infrared Deep Sky Survey (Lawrence et
al. 2007); and in near-ultraviolet and far-ultraviolet filters
from the Galaxy Evolution Explorer (Martin et al. 2005).
We have used the photometry to estimate photometric 
redshifts using the artificial neural network code ANNZ (Smith
et al. 2011).

Our base sample 
for the analysis in this paper is the 
2250 Herschel sources detected at $>5\sigma$ at 250 $\mu$m in the fields common
to H-ATLAS and GAMA that have reliable optical counterparts with 
spectroscopic redshifts $<0.1$.
The fraction of Herschel sources for which we can find counterparts
using the likelihood technique declines with the redshift of the
source, but at $z<0.1$ the fraction of sources with identified
counterparts should be close to 100\% (Smith et al. 2011).
Virtually all the counterparts below this redshift have spectroscopic
redshifts (Smith et al. 2011).

We have estimated the key properties of each galaxy, in particular its star-formation rate,
from the galaxy's spectral energy distribution---the 11 photometric
measurements listed above plus the five measurements in the Herschel bands---using
the galaxy model MAGPHYS (Da Cunha, Charlot
and Elbaz 2008). Some of
the characteristics of MAGPHYS important for our analysis are that it takes account of both the
star formation that is seen directly in the optical waveband
and the star formation that is hidden by dust,
and that the model spectral energy
distribution is fit to the observed spectral energy distribution using a probabilistic method, which
explores a very large number of possible star-formation histories and leads to a probability
distribution for each galaxy property. Further details of the application of
MAGPHYS to the H-ATLAS galaxies are given in Smith et al. (2012b). We use the median estimate of the
star-formation rate for each galaxy, $SFR_i$, derived from the probability distribution of the 
star-formation
rate produced by MAGPHYS.

Our estimate of the star-formation rate per unit comoving volume
in the Universe today is then

\smallskip
\begin{align}
SFR\ density = \sum_i {SFR_i \over V_i}
\end{align}
\smallskip

\noindent in which $V_i$ is the accessible volume in which the galaxy could have been detected in the HATLAS
survey. In calculating the accessible volume, we assume that the spectral energy distribution of the
galaxy in the Herschel bands follows a modified black body, with the dust temperature
and emissivity index both a linear function of the logarithm of the 250-$\mu$m monochromatic
luminosity (Eales et al. in preparation). However, at such low redshifts 
our estimate of the accessible volume of a galaxy depends only very weakly
on the assumption made about its spectral energy distribution.

The GAMA team has measured the value of the S\'ersic index for all these galaxies
from images in nine different passbands (Kelvin et al. 2012). They find that there is a small increase
in the measured value of the S\'ersic index when the wavelength of the image used
to measure the S\'ersic index is increased.
We have used the values they have measured from the
SDSS r-band images,
since this band corresponds well to the typical rest-frame wavelength
of the images used to estimate the S\'ersic indices of the high-redshift galaxies (\S 3.5).
We have measured how the star-formation rate per unit comoving volume in the
Universe today depends on the S\'ersic index of the galaxy in which the stars are forming 
by restricting the 
sum in equation 2 to the galaxies that fall in each bin of S\'ersic index (Fig. 1).
The figure shows that today the star-formation rate per unit comoving volume peaks
at a S\'ersic index of $\simeq$1, implying that, today in the Universe, most stars form in galaxies that
are dominated by a disk.

\begin{figure}
\includegraphics[width=84mm]{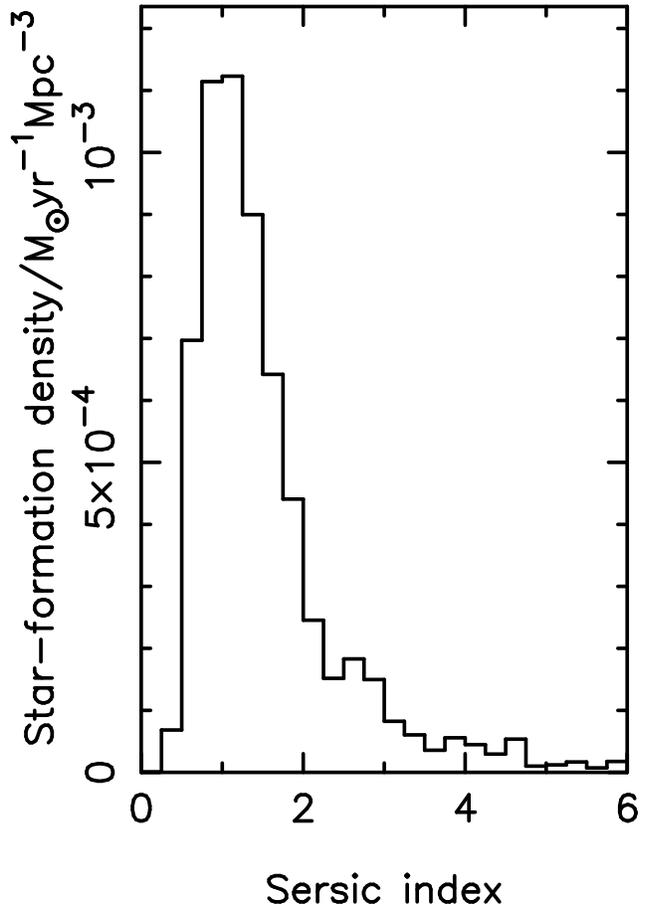}
  \caption{
Star-formation rate per unit comoving volume in the Universe today as a
function of S\'ersic index. 
}
\end{figure}

\subsection{The relation between stellar mass-density and morphology}

We can estimate the stellar mass-density as a function of S\'ersic index in a similar
way. This time we have started from the optical GAMA survey, since 
there are many ETGs that contain large stellar masses
but which are forming stars at a slow rate, have little dust, and are difficult
to detect with Herschel. 
We have started from the GAMA I database, which includes galaxies with
r-band magnitudes brighter than 19.4 in the GAMA9 and GAMA15 fields
and 19.8 in the GAMA15 fields. We exclude galaxies with $z<0.002$, since
their velocities are likely to have a significant component
from peculiar motions. Our sample is the 15724 galaxies with spectroscopic
redshifts in the redshift range $0.002 < z < 0.1$.
The stellar mass-density in the Universe today is given by

\smallskip
\begin{align}
Stellar\ mass\ density = \sum_i {M_{*i} \over V_i}
\end{align}
\smallskip

\noindent in which $M_{*i}$ is the stellar mass of the i'th
galaxy and $V_i$ is again accessible volume. 
We have used the stellar masses for the galaxies derived by Taylor et al.
(2011), since there are not yet MAGPHYS estimates of the stellar mass
for all the GAMA galaxies. 
For the H-ATLAS galaxies, 
for which there are estimates from both methods,
there is a systematic offset of 0.17 in $log_{10}(stellar\ mass)$ between
the two estimates, but once this offset is corrected for, the agreement
is good, with a root-mean-squared discrepancy of 0.13 in $log_{10}(stellar\ mass)$
between the two estimates (Eales et al. 2015). The offset is
not important for this analysis in this section.
We have calculated the accessible volume for each galaxy using an optical spectral
energy distribution derived for each galaxy from its multi-band photometry
(Taylor et al. 2011).
We can estimate the dependence of stellar mass density
on S\'ersic index by restricting the
sum in equation 3 to the galaxies that fall in each bin of S\'ersic index.
Figure 2 shows the stellar mass-density in the Universe today as a function
of S\'ersic index.

As we discussed in \S 1, there is no agreement about the best way to
separate
ETGs and LTGs.
In this paper, we use the operational definition
that ETGs and LTGs are galaxies with a S\'ersic index
greater than and less than 2.5, respectively - a definition
which has been used often in the literature
(e.g. Shen et 
al. 2003; Barden et al. 2005; Buitrago et al. 2013). 
One argument suggests that it is a definition which
has some physical meaning.
Figure 1 shows that the distribution of star-formation rate per unit comoving
volume is fairly symmetric, with a peak at $n=1$.
This value of $n$ is the one expected if the structure of the galaxy
is dominated by an exponential disk, suggesting that the
physical meaning of the peak is that star formation mostly occurs in
disk-dominated galaxies.
Therefore it seems reasonable to devise an ETG/LTG classification that
puts this peak in one class.
With a dividing line of $n=2.5$, the galaxies
in this peak all fall in the class of LTGs, but with a lower value of
$n$ some of the galaxies in this peak would be classified as ETGs;
and with a higher value of $n$ some of the galaxies that are outside this peak
would be classified as LTGs. 
Nevertheless, despite this argument, our definition is basically
an operational one that allows us to compare the morphologies of galaxies
in the Universe today with those of galaxies in the distant past.
Using this definition, we calculate from
the distribution shown in Figure 1 that $\simeq$89\% of the star formation in the
Universe today is contained in LTGs. 

Using this definition, we calculate from Figure 2 that 51\% of the
stellar mass-density today is in ETGs. Kelvin et al. (2014a) have used the optical
images of galaxies in the GAMA survey to classify the galaxies into different Hubble types.
Kelvin et al. (2014b) then used these classifications to show that
34\% of the stellar mass-density today is in elliptical galaxies and
37\% is in S0-Sa galaxies, and so 71\% of the stellar mass-density is
in galaxies that would be traditionally assigned to the ETG class. 
The fact that our estimate is lower is most likely explained by the fact
that
many of the Sa galaxies and and some S0s have 
structures that are dominated by disks (Laurikainen et al.
2010).

Another way to classify galaxies
is to  separate them using their colours into star-forming galaxies
and `passive' or `quiescent' galaxies in which star formation has largely
stopped (e.g. Peng et al. 2010). Although we present arguments
in our accompanying paper (Eales et al. 2015)
for why this classification is misleading and hides some of the physics
that is going on in the galaxy population, it is true
that there is a large overlap between the red passive quiescent galaxies
and ETGs and between the blue star-forming galaxies and LTGs.
Using the stellar mass functions for star-forming and passive galaxies
given in Baldry et al. (2012), we calculate that 73\% of the stellar mass-density
today is in passive galaxies. 
The obvious explanation of our lower estimate of 51\% for
the fraction of the stellar mass-density that is contained
in ETGs is if there is a population
of passive galaxies which have structures that are dominated by disks
(Van den Bergh 1976; Cortese 2012).

\begin{figure}
\includegraphics[width=84mm]{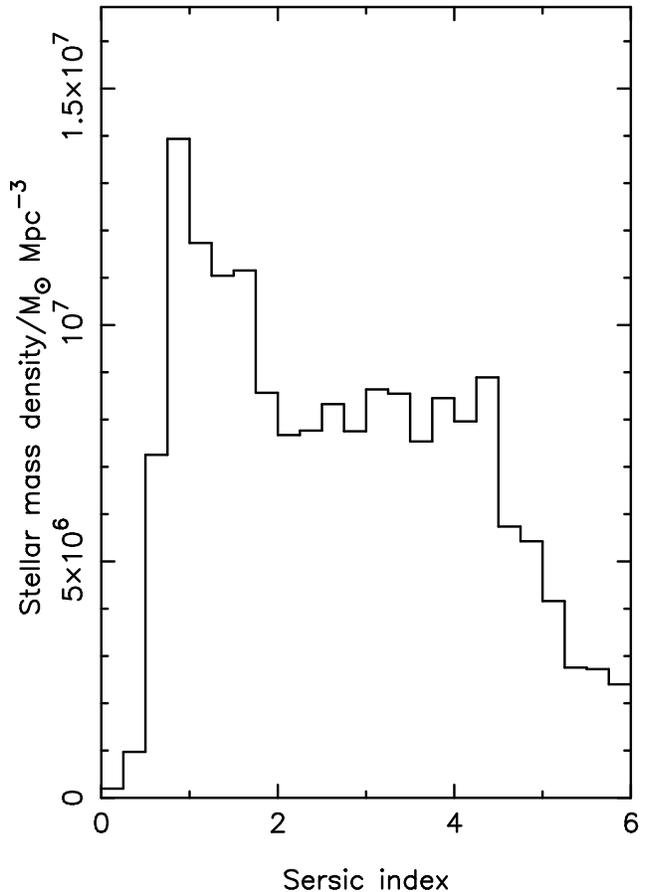}
  \caption{
Stellar mass per unit comoving volume in the Universe today as a
function of S\'ersic index. 
}
\end{figure}

\section{STAR FORMATION OVER THE HISTORY OF THE UNIVERSE - METHOD}

\subsection{The calorimetry equation}

We make the assumption that all the extragalactic background radiation is due to nuclear fusion
in stars. A number of theoretical arguments and observational results show that $\simeq$10\% of the
background radiation comes from the release of gravitational energy in active galactic nuclei
(Dwek and Krennrich 2013), but this should have very 
little effect on our final results since the absolute value of the
intensity of the background is not used in our method.

The basic equation we use links the background radiation, $B_{\nu}$, produced
as the result of nuclear fusion in stars, with the elements that are produced
by the nuclear fusion:

\smallskip
\begin{align}
\int^{\infty}_0 B_{\nu} d \nu = \int^{t_h}_0 {0.007 \rho c^3 \over 4 \pi (1+z)} dt
\end{align}
\smallskip

\noindent In this equation $\rho=\rho(z)$ is
the average co-moving density of elements produced by nuclear fusion
in stars per unit time at a redshift $z$, $t$ is cosmic time, $t_H$ is the age of the Universe and the
numerical factor, 0.007, 
is the fraction of the mass of hydrogen that is turned into energy when hydrogen
is fused into helium. Since the products of nuclear fusion accumulate over time, this
equation allows us to relate the energy dumped into the background radiation by nuclear fusion
at a redshift $z$
to the products of this nuclear fusion in the Universe today.
The calorimetry equation is exact and can be derived directly from the Robertson-Walker
metric, and therefore does not require any assumptions about the values of the standard
cosmological parameters.

Although the equation is exact, for it to be a useful practical tool it is necessary to
make some
approximations. One that has been used in the past (Pagel 1997) 
is to assume that the
density today of all the elements produced by nucleosynthesis, which includes
helium, is proportional to the density today
of the elements heavier than helium (the metals). In
this paper, we have tried the different approach of connecting the
stellar mass-density in the Universe today with the energy produced
from nucleosynthesis in these stars that is now in the background
radiation.

With a few assumptions, which we will discuss below, we can write
the following two equations:

\smallskip
\begin{align}
\rho_* = (1-R) \int^{t_h}_0 SFR(t) dt 
 \end{align}
\smallskip

\begin{align}
 \int^{t_h}_0 SFR(t) dt 
= \int^{\infty}_0 \int^{\infty}_0 {4 \pi (1+z) J(z,\nu) \over
0.007 c^3} d\nu dz
 \end{align}
\smallskip
\noindent in which $\rho_*$ is the current average mass-density of stars
in the Universe, $SFR(t)$ is the mass of stars produced per unit volume per unit time
at a redshift $z$, 
$R$ is the fraction of that newly formed stellar mass that is subsequently ejected
into the interstellar medium as the result of stellar winds
(e.g. Leitner and Kravtsov 2011),
$J(z,\nu)$ is the background radiation produced by objects at
redshift $z$ in a unit interval of redshift, and $t_h$ is again the current age of the
Universe. 

A basic assumption that we have made to produce the first
equation is that
the stars formed in the Universe at a time $t$ are still in the Universe
today.
This assumption seems a reasonable one since most of the mass in a stellar
population is in long-lived low-mass stars.

An assumption made to produce 
the
second equality is that all the
available nuclear energy in stars has been liberated and dumped into
the background radiation by the current epoch, which is obviously not
true because the Sun is still shining. However, it is not a crazy assumption
because most of the energy 
produced by a population of stars is produced by the high-mass stars, which do
burn through their available nuclear energy very quickly.
This is confirmed by Figure 3, where we
have used models of stellar populations (Maraston 2006) to predict how the total energy
that has been emitted by a single stellar population depends on age
for various assumptions about the initial
mass function (IMF). 
The figure shows that for a Salpeter or Kroupa IMF most of the energy has been emitted
by 1 Gyr and the amount of energy that has been radiated (and 
dumped into the background radiation)
increases by
only $\simeq$50\% as the population ages from 1 Gyr to 10 Gyr.

Our main results in this paper, however, were not obtained by assuming either
equality but by combining the two equations in the form of a proportionality
in which we assume that the current stellar mass-density in a class of galaxies,
$\rho_{*c}$,
is proportional to the contribution of the galaxies
in that class to the background radiation:

\smallskip
\begin{align}
\rho_{*c} \propto \left[ \int^{\infty}_0 \int^{\infty}_0 {4 \pi (1+z) J(z,\nu) \over
0.007 c^3} d\nu dz\right]_c
 \end{align}
\smallskip

\noindent The assumptions behind this proportionality are slightly different
from the assumptions behind the equalities.
We are now assuming that the fraction of newly formed stellar
mass lost by stellar winds 
and that the fraction of the available nuclear energy
liberated by the current epoch (see above) do not vary between
galaxy classes. The second quantity is probably different
between ETGs and LTGs, but we argue in \S 5 that the difference
actually strengthens the main result in this paper. 

One further assumption that is built-in to the calorimetric method
is that the IMF does not vary significantly between galaxy classes.
In Figure 3, we have explored the importance of this assumption
by predicting the relationship between bolometric luminosity and
time for different assumptions about the IMF. The figure shows
that the difference between a Kroupa and a Salpeter IMF
is quite small, but our results would be changed significantly
if some galaxy class had a bottom-heavy IMF,
because these galaxies might then contain a large mass
of stars today without having contributed much radiation
to the extragalactic background radiation.

\begin{figure}
\includegraphics[width=84mm]{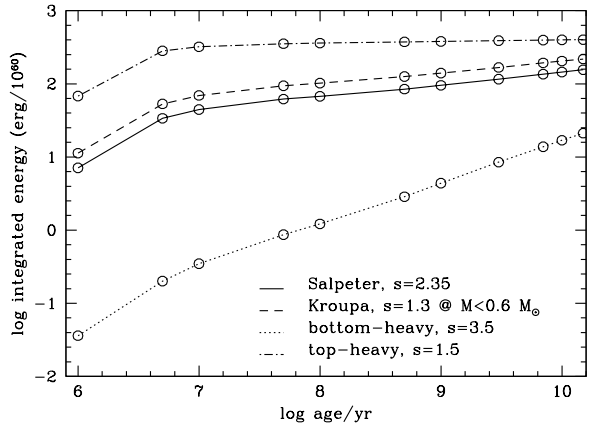}
  \caption{The total energy that has been emitted by a given age for a single stellar population
with a mass of  
10$^{11}$ $M_{\odot}$.
We have used the models of stellar populations of Maraston (2006). The 
different lines are for
different assumptions about the stellar initial mass function. `Kroupa' and `Salpeter' refer to
initial mass functions measured in the solar neighborhood, whereas `top-heavy and `bottom-heavy
refer to initial mass functions with a larger proportion of high-mass and low-mass stars,
respectively.
}
\end{figure}

\subsection{The extragalactic background radiation}

The extragalactic background radiation (Dole et al. 2006; Dwek and
Krennrich 2013) is dominated by the radiation in two wavebands
of roughly equal strength: in the optical/near-infrared, where the background peaks at $\simeq$1-2 $\mu$m ,
and in the far infrared, where the background peaks at $\simeq$100-200 $\mu$m.

The far-infrared background radiation has been measured by the FIRAS and DIRBE
experiments on the Cosmic Background Explorer satellite (COBE) (Fixsen et al.
1998; Dole et al. 2006; Fig. 4). In calculating the
contribution of individual sources to the background (\S 3.3), 
at 160, 250, 350 and 500 $\mu$m 
we used the FIRAS observations of the background. 
The 1-$\sigma$ uncertainty on
the FIRAS estimate of the background is $\simeq$30\%.
At 100 $\mu$m we used the DIRBE
measurement placed on the FIRAS flux
scale (Dole et al. 2006), since the FIRAS measurements did not extend down to this wavelength. 

In the optical/near-infrared, there have been three different techniques
used to estimate the strength of the extragalactic background radiation:
1) estimates from integrating the galaxy counts; (2) absolute measurements of the
background using the DIRBE instrument on COBE (Cambresy
et al. 2001; Levenson et al.
2007); (3) searches for absorption effects in
the gamma-ray spectra of distant active galactic nuclei (Dwek and
Krennrich 2013). The second 
method would be ideal
except for the systematic effects on the DIRBE near-infrared measurements from sunlight scatted
by interplanetary dust. Even after a correction has been made for this scattered light, there is
generally a clear relationship between the measurements of the background radiation and ecliptic
latitude (Cambresy et al. 2001; Levenson et al.
2007), implying a residual contamination from interplanetary dust (reliable
determinations will probably require measurements from beyond 3 A.U. from the Sun, where the
contamination from scatted light is small (Levenson et
al. 2007)). These absolute measurements are 2-4 times
greater than the estimate from integrating the galaxy counts, but upper limits from the gamma-ray
spectra of active nuclei are generally lower than these absolute measurements, also
suggesting that the DIRBE measurements are too high (Dwek and Krennrich 2013). For these
reasons we have taken as
our estimate of the background at 1.6 $\mu$m the integral of the deep 1.6-$\mu$m ($H_{AB}$) galaxy counts
made with Wide Field Camera 3 on the Hubble Space Telescope (Windhorst et al.
2011). 

\subsection{Samples of the background radiation}

To apply the calorimetry technique (\S 3.1), we ideally need to be able to resolve the extragalactic
background radiation into individual sources. By measuring the structures
of the galaxies producing the emission, we can then use equation 7 
to determine the proportion
of stars that have formed in galaxies with different structures.

In the southern field of the Great Observatories Origins Deep Survey (GOODS-South),
we can get surprisingly close to this ideal. This field has been surveyed by the deepest survey with
the Herschel Space Observatory,
Herschel-GOODS (Elbaz et al. 2011; Magnelli et
al. 2013) at 100 and 160 $\mu$m, close to one of the two peaks in the extragalactic
background radiation. It has also been observed as part of the  
Cosmic Assembly Near-Infrared Deep Extragalactic Legacy Survey
(CANDELS), an optical/near-infrared survey carried out with the Hubble Space Telescope 
with a main wavelength of 1.6 $\mu$m, close to
the other peak of the extragalactic background
radiation (Koekemoer et al. 2011;
Guo et al. 2013). The CANDELS team has measured S\'ersic indices
for the galaxies detected in the survey (Van der Wel et al. 2012) from data 
that are well matched to the ground-based
images used to estimate the S\'ersic indices in our investigation of the Universe
today (\S 2, \S 3.5).
We now show how these surveys can be used to resolve the two peaks in the extragalactic
background radiation into individual sources.
Although the total area covered by the surveys of GOODS-South is only $\simeq$100 arcmin$^2$,
we show in \S 4 that cosmic variance is not a significant problem, and thus that the
samples of sources provided by these surveys constitute fair samples of the
extragalactic background radiation.

\subsubsection{Sampling the far-infrared background with the sample of Herschel sources}

The public catalogue for Herschel-GOODS (H-GOODS) is available at http://hedam.lam.fr/GOODS-Herschel/.
This catalogue was obtained by starting with the assumption that the sources in the
100 and 160-$\mu$m images are at a set of `prior positions' drawn from a catalogue of sources
detected at 24 $\mu$m with the Spitzer telescope, and then varying the 100-$\mu$m and 160-$\mu$m flux
densities of the sources to match the observed structure in the images (Elbaz et al.
2011; Magnelli et al. 2013). The Spitzer
catalogue itself was obtained in a similar way from a set of prior positions drawn from a
catalogue of sources detected at 3.5 $\mu$m with the Spitzer telescope (Elbaz et al. 2011;
Magnelli et al. 2013). This technique has
the advantage that it produces very accurate positions for the H-GOODS sources because the
3.5-$\mu$m sources have much more accurate positions than is possible to obtain with Herschel
observations by themselves. However, it has the potential disadvantage that it could miss
Herschel sources if they are not present in the prior catalogue. By comparing the H-GOODS
catalogue with one obtained without using the prior catalogue, the H-GOODS team estimates
that the catalogue made using the prior positions misses $<$4\% of the sources at 160 $\mu$m (Magnelli et
al. 2013).
Furthermore, any missing sources will be among the fainter ones, and thus should have a
negligible effect on the calorimetric technique.

The H-GOODS catalogue contains sources that were detected at 100 and 160 $\mu$m at $>3\sigma$.
Sample 1 in our analysis contains the 527 sources detected at $>3\sigma$ at 160 $\mu$m in the region of
GOODS-South also covered by the CANDELS survey. 
To estimate the contribution of this  sample to the background radiation, we need to
make measurements of the flux density at 100, 160, 250, 350 and 500 $\mu$m for every source in
the sample; even if an individual detection is below the 3$\sigma$ level, the result of summing the
results for all the sources is often highly significant.
We essentially carry out a
statistical `stacking analysis' to estimate the contribution from
all the sources, even those that are not detected
individually at a significant level (Marsden et al. 2009).

There are no H-GOODS images at the three longer wavelengths, so at these wavelengths
we measured
the flux densities of the sources from images obtained as part of the HERMES
survey (Oliver et al. 2012). To 
measure the flux density of a source, the standard technique is to first convolve
the image with the point spread function of the telescope, since this maximizes the 
signal-to-noise
of the source. However, this is only correct if the noise on the image is uncorrelated
between pixels, which is true of the instrumental noise on the HERMES images. Most of the
noise on the HERMES images, however, is confusion noise, the noise due to faint submillimetre
sources that are too faint to detect individually, and this noise is correlated between pixels. We
did not therefore convolve the HERMES images but simply measured the flux densities from the
pixel in each HERMES image that contained the position of the H-GOODS source. 

The background level on a Herschel image is usually close to but not exactly zero.
To estimate the residual background emission and the errors on the flux densities,
we
used the
Monte-Carlo approach of measuring flux
densities at a large number of random positions on the HERMES images without
making a correction for any residual background emission. 
We found background levels of
-0.48$\pm$0.08 mJy, -0.74$\pm$0.08 mJy and -1.1$\pm$0.08 mJy at 250, 350 and 500 $\mu$m, respectively,
which we used to correct our measurements of the 
flux densities of the H-GOODS sources. From the Monte-Carlo simulation, the
errors on these flux densities
at 250, 350 and 500 $\mu$m are 6.2 mJy, 6.5 mJy and 6.2 mJy, respectively, very
similar to the Herschel confusion noise at these wavelengths (Nguyen et
al. 2010). 

We used the 160-$\mu$m flux densities from the catalogue,
but there was often not a 100-$\mu$m measurement in the catalogue.
If a measurement did not exist in the catalogue, we used aperture photometry to
measure the flux density from the H-GOODS 100-$\mu$m image, using 
an aperture with a radius equal to the
full-width-half-maximum of the point spread function (6.7 arcsec).
We again used a Monte-Carlo
approach to estimate the residual background level on the image and the errors on the
flux densities, 
performing aperture photometry at a large number of random positions on the 
image without correcting for a residual background. 
We found a residual background level in our aperture of 
0.3$\pm$0.03 mJy, which we used to correct the aperture photometry
of the H-GOODS sources.
For the H-GOODS sources for which 100-$\mu$m photometry exists in the public catalogue,
we compared our measured flux densities with the catalogued flux densities to derive an aperture
correction to bring our measurements on to the same flux scale as the catalogue. As a result of
this comparison, we have multiplied our aperture flux densities by a factor of 1.82.
From the Monte-Carlo simulation, the error on
our measured flux densities is 2.4 mJy.

We estimated the contribution of the sample to the background radiation at each
wavelength by adding the flux densities of the individual sources. We estimated the error on the
sum by adding in quadrature the errors on the individual flux densities. 
The errors on the total flux density of the sources are 2\%, 0.6\%, 4\%, 5\% and 10\% at
100, 160, 250, 350 and 500 $\mu$m, respectively, much less than the errors on the measurements
of the background radiation at these wavelengths.
Thus the errors on the percentage of the
background radiation resolved by the sources are dominated by the
errors on the background measurements themselves, which we assume are 30\% for
the measurements with FIRAS at the four longest wavelengths and 44\% for the
DIRBE measurement at 100 $\mu$m (Dole et al. 2006). Using these
errors, our estimate of the fraction of the background contributed
by the H-GOODS sample is $0.55^{+0.24}_{-0.17}$ at 100 $\mu$m,
$0.67^{+0.22}_{-0.15}$ at 160 $\mu$m, 
$0.48^{+0.14}_{-0.11}$ at 250 $\mu$m,
$0.50^{+0.15}_{-0.12}$ at 350 $\mu$m,
and $0.45^{+0.14}_{-0.10}$ at 500 $\mu$m (Fig. 4).
The H-GOODS team has estimated the fraction of
the background resolved by the H-GOODS sources as 75\% at both 100 and 160 $\mu$m (Magnelli et al.
2013). There is therefore
a sizeable 
difference at 100 $\mu$m for which we do not know the reason.

The sample of sources detected with Herschel in the H-GOODS survey
therefore constitutes a significant fraction of the background radiation but still falls significantly
short, whether we use our estimates or the estimates of the H-GOODS team, from resolving all the
background radiation.

\subsubsection{Sampling the far-infrared background with a sample
of Spitzer sources}

We show now that we can
actually sample a larger fraction
of the far-infrared background
using a sample of sources detected at 24 $\mu$m in the
GOODS-South region by the
Spitzer Space Telescope. This sample (sample 2) consists of 1557 sources detected at 24 $\mu$m in the
same area of GOODS-South that is covered by CANDELS.
We measured the flux densities of each of these sources at 250, 350 and 500 $\mu$m from the HERMES
images in the same way as for sample 1.
At 100 and 160 $\mu$m, we used for preference the flux densities given in the H-GOODS public
catalogue. If there was not a 100-$\mu$m measurement in the public catalogue, we carried out
aperture photometry on the H-GOODS 100-$\mu$m image in exactly the same way as for sample 1. For 
sample 2
this was often also necessary at 160 $\mu$m.
In this case we measured a 160-$\mu$m flux density from the H-GOODS
160-$\mu$m image through an aperture with a radius equal to the
full-width-half-maximum of the point spread
function (11.0 arcsec). 
We used the Monte-Carlo method of performing aperture
photometry at a large number of random positions on the H-GOODS 160-$\mu$m image
to estimate the residual background level in
this aperture of 2.0$\pm$0.07 mJy, which we allowed for in the measurements of
the flux densities of the H-GOODS sources.
By comparing our measured flux densities
with the ones in the catalogue, where they exist, we derived a correction factor of
2.0, by which we multiplied our aperture flux densities to bring them on to the
same flux scale as the catalogue. From the Monte-Carlo simulation, our
estimate of the error
in our 160-$\mu$m flux densities
is 5.6 mJy.

For sample 2, the error in the total flux density of the sources is 5\%,
7\%, 4\%, 5\% and 8\% at 100, 160, 250, 350 and 500 $\mu$m, respectively, which
are again much less than the errors in the level of the backround.
Using 30\% as the error in the FIRAS measurements of the background
at the four longest wavelengths and 44\% as the error in the DIRBE
background measurement at 100 $\mu$m, we calculate that
the sources in sample 2 compose
$0.67^{+0.29}_{-0.20}$ of the background radiation
at 100 $\mu$m,
$0.90^{+0.27}_{-0.21}$ at 160 $\mu$m,
$0.80^{+0.24}_{-0.19}$ at 250 $\mu$m,
$0.89^{+0.27}_{-0.21}$ at 350 $\mu$m,
and $0.90^{+0.27}_{-0.21}$ at 500 $\mu$m.
Given the uncertainties, the combined far-infrared flux density from
sample 2 is fully consistent with the far-infrared background
radiation over this range of wavelengths.
The sources in this sample therefore 
represent a fair sample
of the extragalactic background radiation in one of its two main
peaks.

\begin{figure}
\includegraphics[width=84mm]{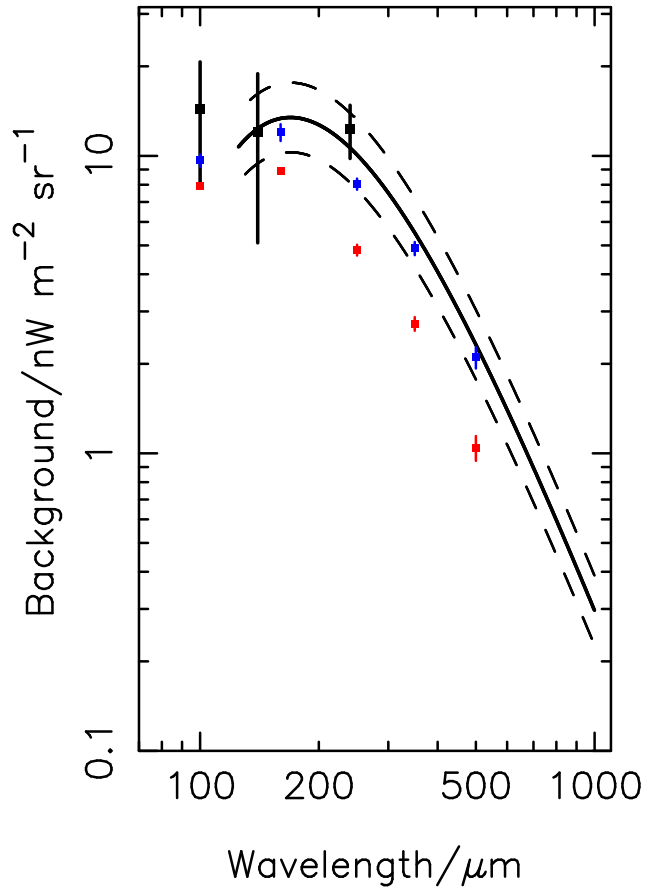}
  \caption{The extragalactic far-infrared background radiation plotted against 
wavelength, and our
estimates of the contribution to this background from the samples described in this paper. The
heavy solid line shows the measurement of the background with the FIRAS experiment on
COBE (Fixsen et al. 1998), with the light dashed lines showing the $\pm1\sigma$ errors on 
the background. The black
points show the measurements made by the DIRBE experiment on COBE (Dole et al.
2006). The red points
show our measurements of the contribution to the background radiation from the sources
individually detected at 160 $\mu$m by the H-GOODS survey (sample 1). The blue points show our
measurements, using the Herschel images of GOODS South, of the contribution to the
background radiation from the sources that were originally detected at 24 $\mu$m with Spitzer
(sample 2).
}
\end{figure}

\subsubsection{Sampling the optical/near-infrared background with the HST}

The other peak in the extragalactic background radiation is in the optical/near-infrared
waveband. As our sample of this peak (sample 3) we use the 8488 galaxies detected
by CANDELS with $H_{AB} < 24.5$ in the GOODS-South field. We have used 
this photometric band because its central wavelength ($\simeq$1.6 $\mu$m) is close to
the peak of the optical/near-infrared background. 
We have used the magnitude limit because galaxies fainter than this
do not generally have reliable measurements
of the S\'ersic index (Van der Wel et al. 2012).
This sample constitutes $\simeq$94\% of the extragalactic background radiation
at 1.6 $\mu$m. The high value is the result of our assumption that the extragalactic
background radiation at this wavelength is given by the integral of the galaxy counts (\S 3.2);
since the slope of the galaxy counts flattens at $H_{AB} \simeq 21$ (Windhorst et
al. 2011), the fraction of the background radiation produced by galaxies
with $H_{AB} < 24.5$ is very high.
As long as there is not a substantial contribution to the extragalactic background
radiation from starlight that is not detected in even the deepest exposures with
the HST (\S 5), sample 3 constitutes a fair sample of the extragalactic background radiation
in its second peak.

We 
have included all the data for samples 1 and 
2 that was used in this paper in on-line tables.
The tables includes the positions of the Herschel and Spitzer sources, the flux densities at
the five Herschel wavelengths, the spectroscopic and/or photometric
redshifts of the sources, the positions of the counterparts on the HST images
(\S 3.4) and the S\'ersic indices of the counterparts (\S 3.5).

\subsection{Identifying the counterparts to the Herschel and Spitzer sources}

There are measurements of the S\'ersic index for all the galaxies in sample 3
by the CANDELS team (Van der Wel et al. 2012). However, for the other two
samples, before we can consider the structures of the galaxies producing the
far-infrared emission, we need to identify these galaxies.

The positions of the sources in samples 1 and 2 are
ultimately based on observations with the Spitzer Space Telescope at 3.5 $\mu$m, which have an angular
resolution (FWHM) of 1.7 arcsec. The accuracy of the positions should therefore be $<$1 
arcsec,
and we have made the assumption that any genuine counterpart 
to a Herschel/Spitzer source
on the
CANDELS H-band (1.6 $\mu$m) image
must lie within 1.5 arcsec of the position of the 
source.

To assess the probability of a possible counterpart on the CANDELS
image within 1.5 arcsec of the position of the Herschel/Spitzer source
being there by
chance, we follow the frequentist technique commonly used in far-infrared astronomy of
calculating a statistic, $S$, for the counterpart, defined as the number of CANDELS galaxies
brighter than the 1.6-$\mu$m flux density of the potential counterpart that are expected by chance to
lie closer to the Herschel/Spitzer position than the potential counterpart (Dye et al.
2009). If there was more than one
possible counterpart within 1.5 arcsec, we chose the one with the lowest value of $S$ as the most
likely counterpart. We estimated the probability, $P$, of obtaining such a low value of $S$, given the
null hypothesis that there is no relation between the CANDELS galaxies and the Herschel/Spitzer
sources, by the Monte-Carlo approach of measuring $S$ for a large number of random positions on
the
image. Sample 1 contains 527 sources and we found possible counterparts
within 1.5 arcsec for 526. We estimated the number of counterparts that are there by chance by
summing $P$ for the 526 counterparts, which is 3.0. Therefore $<$1\% of the counterparts are likely to be
spurious. Sample 2 contains 1557 sources and we found possible counterparts within 1.5 arcsec
for 1537. In this case, the sum of $P$ is 9.9, again showing that $<$1\% of the counterparts are likely
to be spurious.

From a formal statistical point of view, there is no reason why this frequentist technique
should produce any biases. However, one might worry that, because the flux at 1.6 $\mu$m is used in
the calculation, the technique is biased against highly obscured dusty galaxies, the kind that one
might expect to be far-infrared sources. However, an investigation shows that this is not the case.
In sample 1, of the 527 160-$\mu$m sources, there were 92 cases where there was more than one
potential HST counterpart within 1.5 arcsec. In only 5 of the 92 cases was there an object fainter
than the selected counterpart but closer to the Herschel position. A final argument that the
method is reliable is that 77\% of the selected counterparts lie within 
0.2 arcsec of the 
position of the Herschel
source, whereas our Monte-Carlo analysis shows that the distribution of offsets for random
positions reaches a maximum at $>$1 arcsec.

\subsection{The S\'ersic indices of the CANDELS galaxies}

In this section we consider whether any systematic errors might arise from
comparing the S\'ersic indices of the H-ATLAS galaxies (\S 2) with those measured
for the galaxies detected in CANDELS.

The S\'ersic indices for the H-ATLAS sources were measured from the r-band (0.62 $\mu$m)
images from the Sloan Digital Sky Survey (SDSS) (Kelvin et al.
2012). The S\'ersic indices for the H-GOODS
galaxies were measured from H-band (1.6 $\mu$m) images from the CANDELS HST programme
(Van der Wel et al. 2012).
The median redshift of the H-ATLAS galaxies is 0.07, whereas most of the H-GOODS
galaxies lie at $z>1$ (\S 4). For these redshifts, the HST images 
of the H-GOODS galaxies (angular resolution
of 0.17 arcsec) and the SDSS images of the H-ATLAS
galaxies (angular resolution $\simeq$1 arcsec) 
give almost exactly the same physical resolution, $\simeq$1.3 kpc. The r-band image
of a galaxy at $z=0$ and the H-band image of an H-GOODS galaxy at $z=1.6$, which is fairly
representative of the H-GOODS redshift distribution (\S 4), correspond to the same rest-frame
wavelength. The similarity of the rest-frame wavelengths ensures that there should be no biases
created by the small changes that are evident when S\'ersic indices are measured at different
wavelengths (Kelvin et al. 2012). The one slight mismatch is in 
the sensitivity to surface brightness for the two
datasets. The HST image of an H-GOODS galaxy at $z=1.6$ has $\simeq$1 mag less surface-brightness
sensitivity than the SDSS 
image at the same rest-frame wavelength
of a galaxy at $z=0$, and the difference increases
at $z>1.6$. 
Van der Wel et al. (2012) have investigated the effect of the signal-to-noise
on the CANDELS images
on the measured value of the S\'ersic index, concluding that there is little bias
down to $H_{AB}=23.0$ but that by $H_{AB} = 24.0$ galaxies with high values of
the S\'ersic index ($n>3$) will have measured values that are lower than the
true values by 25\% (see their Table 3). Therefore, some of the fainter
objects in our high-redshift sample may have measured S\'ersic indices that are too low.

\subsection{Practical application of the calorimetry technique}

Subject to the assumptions discussed in \S 3.1, equation 7 shows that
the stellar mass-density in the Universe today associated with the star-formation
in the i'th source in a fair sample of the background radiation is
proportional to
$J_i (1+z_i)$, in which $z_i$ is the redshift of the galaxy
and $J_i$ is the contribution of this galaxy to the extragalactic background radiation
\footnote{There is no causal association, of course. We are using the samples
to represent the energy generated by nucleosynthesis over the history of the
Universe, but the starlight we are detecting in the early Universe has no causal
connection with the stars we see today. We are implicitly assuming the cosmological
principle that any region of the Universe looks like any other.}.
Therefore, given estimates of $J_i$ and $z_i$ for all
the galaxies in a fair sample of the background radiation, it is straightforward to
estimate the stellar mass-density in the Universe today
produced as the result of the formation of stars over the history of the Universe.
Given measurements of the S\'ersic index for each galaxy in a sample, it is also straightforward
to estimate how the stellar mass-density today
{\it depends on the S\'ersic index of the galaxy in which those stars were formed.}
It is important to note that these stars may well now be in a galaxy
very unlike that in which the stars were formed, either as the result of the merging
of galaxies or some other process that has transformed the structure of the galaxy.

Since the flux densities of individual
galaxies are often detected with very low signal-to-noise or are even negative, we calculated $J_i$
in both the optical/near-infrared and the far infrared using the following formula, which provides
a robust estimate in this situation:

\smallskip
\begin{align}
J_i = \sum_j F_{ij} \Delta \nu_j
\end{align}
\smallskip

\noindent In this formula $F_{ij}$ is the flux density of the i'th galaxy in the j'th band and 
$\Delta \nu_j$ is the frequency
range associated with that band. We adopted the very simple approach of assuming that the
boundaries between two bands, for example the Herschel bands centred at 100 and 160 $\mu$m, lie
equidistant in log(frequency) between the two central wavelengths.
For samples 1 and 2, we calculated $J_i$ from the measured flux densities of each
galaxy at 100, 160, 250, 350 and 500 $\mu$m. For sample 3, we calculated $J_i$ from
measurements of the flux density in the following photometric bands
(Guo et al. 2013): U-band, ACS F606W, ACS F814W, WFC125W, WFC160W, IRACB1 and IRACB2.

For redshifts we used, in
order of preference, spectroscopic redshifts, the photometric redshifts that form part of the H-GOODS
public data release and the catalogue of photometric redshifts for the CANDELS
galaxies (Li-Ting et al. 2014).

After losing $\simeq$1\% of the sources for which we could not find
a counterpart on the CANDELS image,
samples 1 and 2, contain 
526 and 1537 galaxies, respectively.
We removed $\simeq$10\% of the galaxies in
each sample for which the 
flag for the S\'ersic measurement (Van der Wel et al. 2012) implies the measurement
is not reliable, leaving 467 and 1358 galaxies in samples 1 and 2, respectively. Of the galaxies
that are left, 352 (75\%) in sample 1 and 903 (66\%) in sample 2 have spectroscopic redshifts; for
the remaining galaxies we used the photometric redshifts.

In the optical/near-infrared, sample 3 contains 8488 objects with $H_{AB} < 24.5$. The small
size of the field means that there is the potential for effects due to cosmic variance (\S 4). These
effects will be largest at low redshifts, and indeed there are a large number of galaxies at $z\simeq0.05$.
To minimize these effects we excluded 123 galaxies with redshifts $< 0.06$. We also excluded 15
very bright objects with S\'ersic indices of $<0.6$, which are probably stars or quasars. Of the
remaining objects, we eliminated $\simeq$20\% for which the flag implies there is not a reliable
measurement
of the S\'ersic index (Van der Wel et al. 2012). Of the remaining 6753 galaxies, 2482 (37\%) have spectroscopic
redshifts, and for the others we used the photometric redshifts.

\begin{figure*}
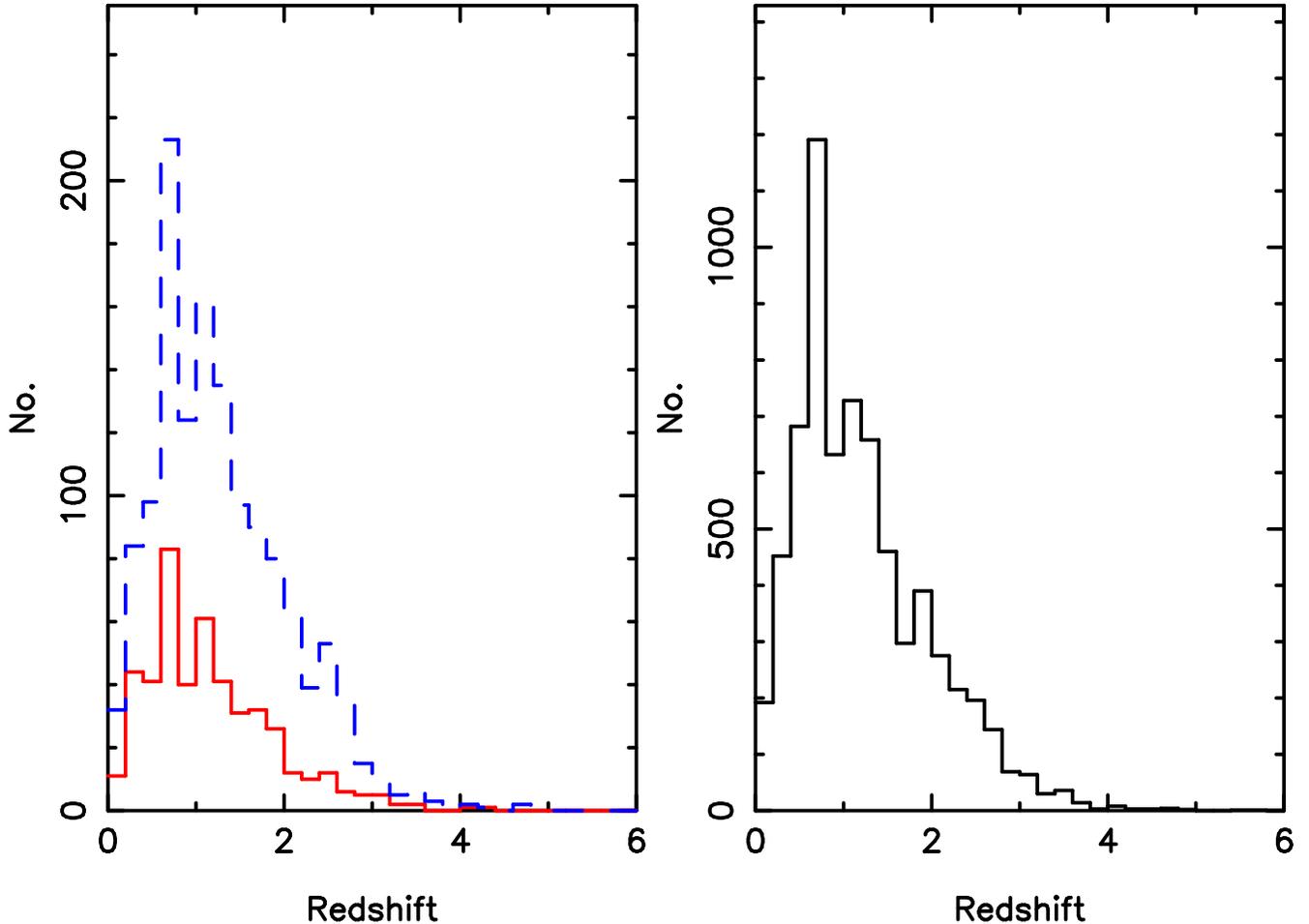

 \centering
 \includegraphics[width=88mm,keepaspectratio]{Fig5a.eps}
 \includegraphics[width=88mm,keepaspectratio]{Fig5b.eps}
 \caption{
Redshift distributions of the samples of the far-infrared background radiation (left)
and the optical/near-infrared background radiation (right). The red line in the left-hand panel shows the
redshift distribution for the sample of sources detected by Herschel (sample 1) and the blue
dashed line shows the redshift distribution for the sample of sources detected with Spitzer
(sample 2).
}
\label{fig:pixhist}
\end{figure*}

\section[]{Results}

Figure 5 shows the redshift distributions of the three samples. Since GOODS-South is such a
small field, an obvious concern is whether it is a fair sample of the Universe and, in particular, of
the extragalactic background radiation. We estimated the cosmic variance for the three samples
using the on-line calculator described by Treventi and Stiavelli
(2008). Using a mean redshift of 1.0 and a redshift interval
of 2.0, we estimate that cosmic variance leads to errors in the total numbers of sources in
samples 1, 2 and 3 of 7, 6 and 5\%, respectively. Therefore, the samples as a whole are 
likely to be fair
samples of the overall background radiation. When we apply the on-line calculator to the
individual redshift bins, which have a width of 0.2 in redshift, we find that the uncertainty in the
number of sources in each redshift bin is $\simeq$30\% for samples
1 and 2 (left-hand panel
of Figure 5). Hence the lumpy nature
of the redshift distributions in Fig. 5 is probably the result of large-scale structure.
Note that there are well-known over-densities in GOODS-South at $z=0.66$ and $z=0.735$
(Adami et al. 2005).

Using the values of $J_i (1+z_i)$ and equation 6,
we have estimated the star-formation history of the
Universe: the star-formation rate per unit comoving volume as a function of redshift.
Figure 6 shows the star-formation history derived from the data for the three samples.
Not surprisingly, at a given redshift the star-formation rate for sample 2
is higher than for sample 1 because the sources in sample 2 comprise a larger fraction
of the background radiation. The star-formation histories
derived from all three samples are rather noisy, the consequence of large-scale structure, but all
three star-formation histories peak somewhere in the redshift range $1<z<3$, consistent with
the results of previous attempts to derive the Universe's star-formation history (e.g. Hopkins 
and Beacom 2006).

\begin{figure*}
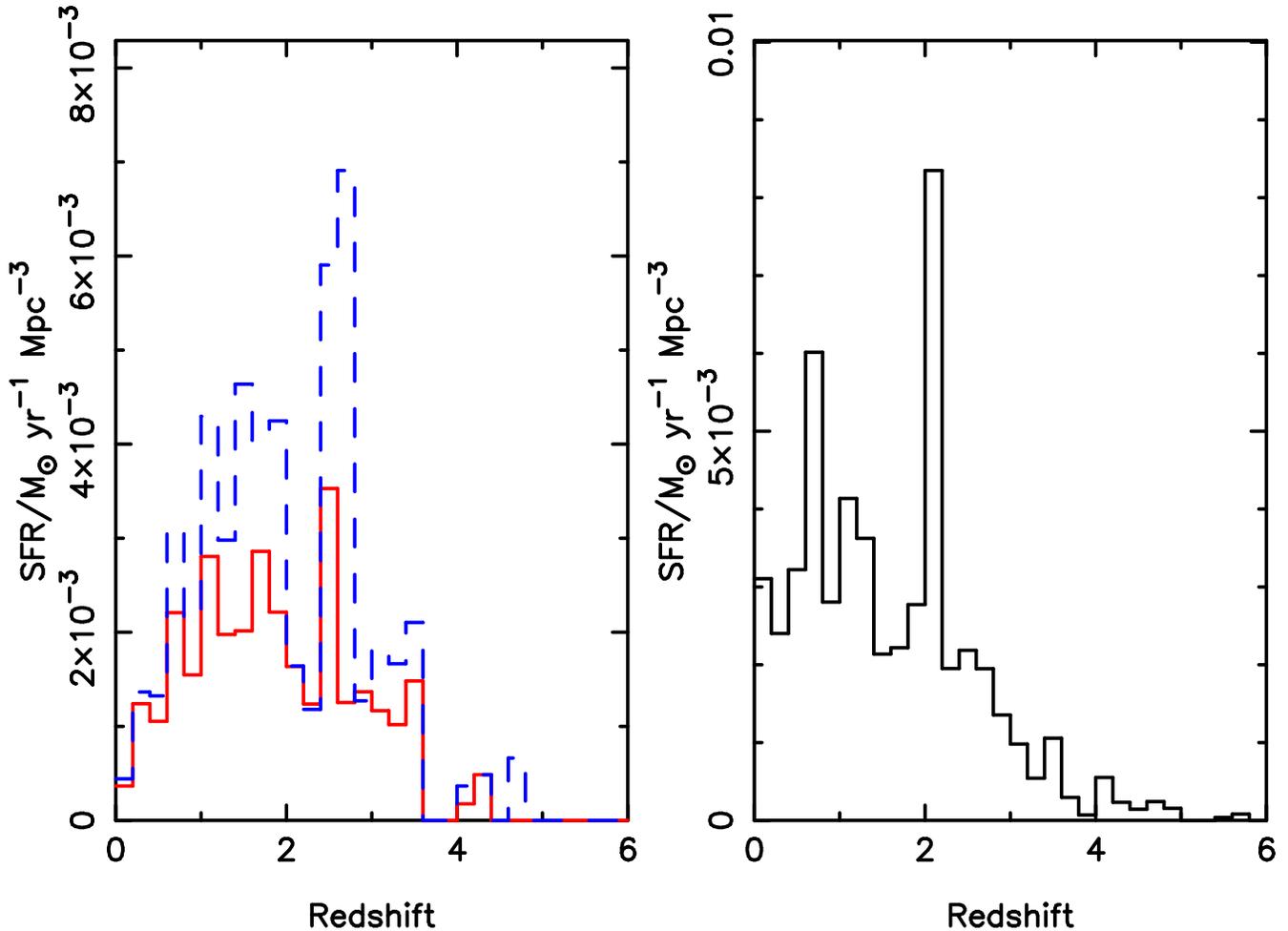

 \centering
 \includegraphics[width=88mm,keepaspectratio]{Fig6a.eps}
 \includegraphics[width=88mm,keepaspectratio]{Fig6b.eps}
 \caption{
Star-formation density (star-formation rate per unit comoving volume) derived by
applying the calorimetric technique to the far-infrared background radiation (left) and to the
optical/near-infrared background (right). In the left-hand panel, the red line shows the results for sample
1 and the blue line the results for sample 2.
}
\label{fig:pixhist}
\end{figure*}

Figure 7 shows the estimates from the three samples for the relationship between the
stellar mass-density in the Universe today and the S\'ersic index of the galaxy in which 
those stars contributing to that mass-density  were formed.
We showed in \S 2 that
in the Universe today 51\% of the stellar mass density is in ETGs,
which we empirically defined as galaxies with a S\'ersic index $>2.5$.
If the structures of these galaxies have not changed between the time the stars now in them were
formed and the current epoch, we would expect 
Figure 7 to look very like Figure 2.
On the contrary, the distributions for all three samples are strikingly similar
to Figure 1, which showed how the star-formation rate in the Universe today
depends on the S\'ersic index of the galaxy in which the those stars are being formed.
Figure 1 showed, not unexpectedly, that, today, most stars form in LTGs.
Figure 7 shows that this has always been true, that even though most stars today are
in ETGs, most stars formed in LTGs.

We can quantify this very easily by measuring the proportions of the distributions in
Figure 7 that lie on either side of the $n=2.5$ dividing line. 
The result from both the samples of the far-infrared background, samples
1 and 2, is that $\simeq$87\% of the
stellar mass that is in the Universe today must have formed in LTGs.
The result from the sample of the optical/near-infrared background, sample 3,
is that
$\simeq$79\% of the stellar mass today was formed in LTGs. 
To assess the importance of the measurement errors for the S\'ersic indices on these
estimates, we carried out a bootstrap analysis for the smallest sample, sample 1.
We generated 1000 artificial versions of sample 1, using a random-number generator
to produce from the measured S\'ersic index and measurement error
for each galaxy
a new `measurement' in each sample for that galaxy.
From these 1000 samples, we estimate the statistical uncertainty in our estimate
of 0.87 for the fraction of the stellar mass-density formed in LTGs as only $1.9 \times
10^{-6}$. We discuss other more important sources of error in the
following section.

Taking an
average of the results for the two peaks in the background
radiation, which is reasonable since they have roughly equal
strength, our results imply that $\simeq$83\% of the stellar mass today was formed in LTGs.
This is much higher than our estimate of 49\% for the percentage of the stellar
mass-density that is today in LTGs. The big difference in the values implies
that the morphological transformation of galaxies must be
an important part of galaxy evolution.

\begin{figure*}
\includegraphics[width=160mm]{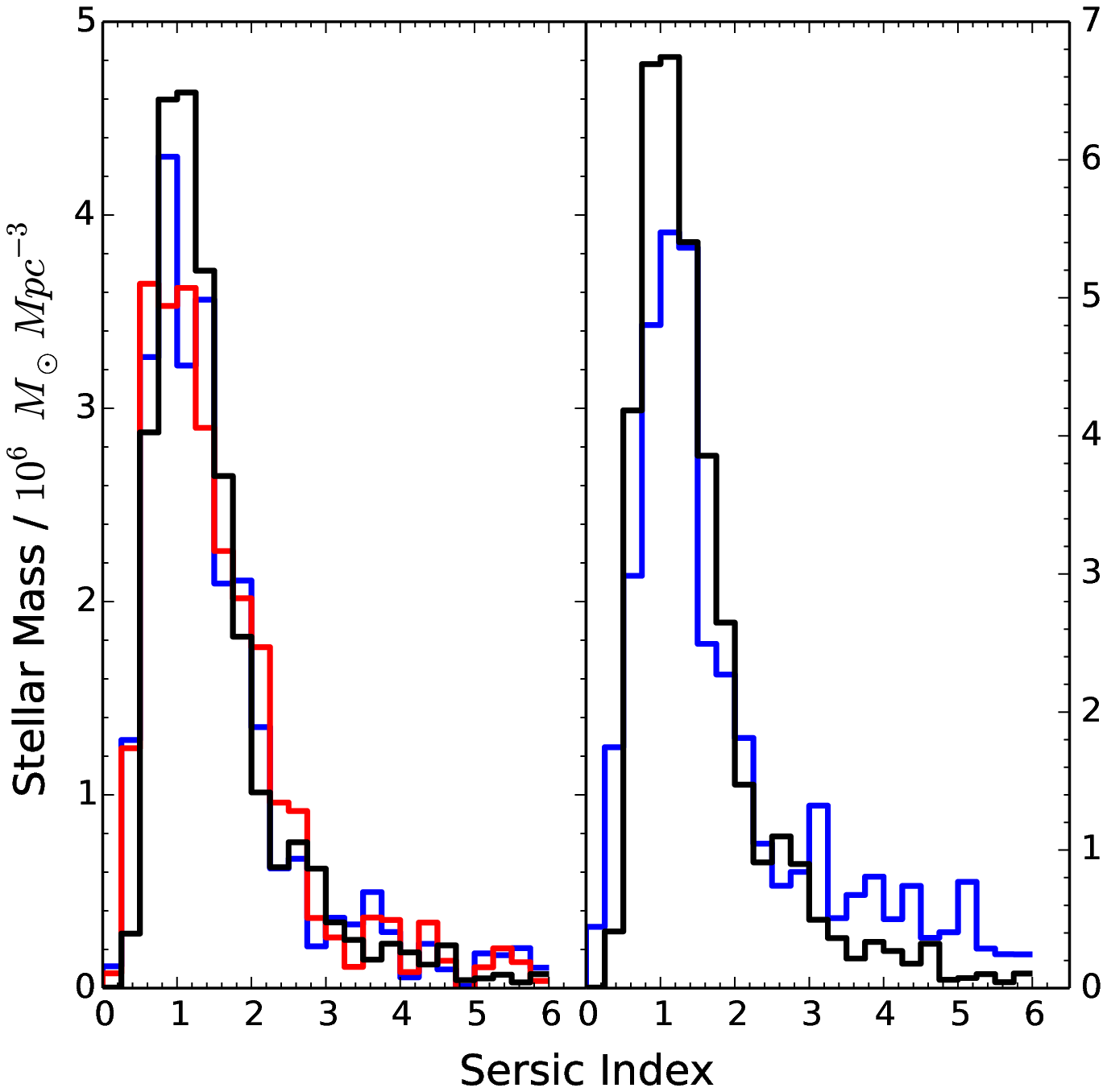}
  \caption{
The relationship between the stellar mass-density in the
Universe today and the S\'ersic index of the
galaxy in which the stars were formed. The two panels show the results for our calorimetric
samples of the far-infrared (left) and optical/near-infrared (right) backgrounds; the blue and red
line in the left-hand panel show the results for samples 1 and 2, respectively, and the blue line in
the right-hand panel shows the results for sample 3. 
The histogram for sample 1 in the left-hand panel has been normalised
so that the area under it is the same as for sample 2.
In both panels the black line shows the star-formation
rate in the Universe today as a function
of S\'ersic index (Fig. 1) normalized so that it encloses the same area as
the other histograms. Note the absence of a peak at $n=4$, which should be present if stars that are
currently in elliptical galaxies were born in galaxies with the same structures.
}
\end{figure*}

\section{Discussion}

The importance of the calorimetric approach is 
that it produces a quantitative measurement of the stellar mass-density that has
formed in galaxies with different morphologies which covers
all cosmic epochs and
galaxy masses, 
since the extragalactic background 
radiation contains all the energy ever emitted as the result
of nuclear fusion in stars. The calorimetric technique, if there are no practical
problems---we will discuss some of the possibilities next---makes it impossible
for star formation to hide anywhere.

One fundamental problem we can do nothing about is that the
method is based on the assumption of a universal initial
mass function. As we showed in \S 3.1, a population of
stars that are formed with a bottom-heavy initial mass
function may lead to a large stellar mass-density in the Universe
today but not have dumped much radiation into the extragalactic
background radiation. 

A less fundamental problem, but one we can still
do nothing about, is that the uncertainties on the absolute values
of the extragalactic background radiation still leave some
places for star formation to hide; if the far-infrared background
is at the upper end of the range set by the errors (Fig. 4), or if the
optical/near-infrared background has a component that is
not detected in the deep HST surveys, our samples may comprise
substantially smaller percentages of the background radiation than
we have estimated. 

Third, we have missed some of the energy produced by nucleosynthesis in stars
because we have only considered the background radiation over two limited ranges
of wavelength around each of the two peaks in the
background radiation (\S 4). This will only be a problem if the galaxies comprising the
missing part of the background radiation have systematically different morphologies
from the galaxies in our samples of the two peaks. This is possible, but seems unlikely
because the two peaks are separated by a factor of $\simeq$100 in wavelength, yet the
morphological makeup of the galaxies in the samples of the
two peaks are quite similar (compare the right and left panels
of Fig. 7).

Fourth, there is the issue that 
for any population of stars, not all the available nuclear energy will
have been dumped into the extragalactic background radiation by the
current epoch (\S 3.1, \S 4). 
ETGs
generally have older stellar
populations than LTGs (Kennicutt
1998) 
and thus should have
dumped a larger fraction of the total available nuclear energy into the background
radiation by the current epoch, and so if there is a systematic bias it will be
in the sense that
samples of the extragalactic background radiation preferentially
contain ETGs. Therefore, this may have led us to 
underestimate the true
difference beween the fraction of the stellar mass-density
that is in LTGs today and the fraction of the total stellar mass-density
that was formed in LTGs.

\begin{figure}
\includegraphics[width=84mm]{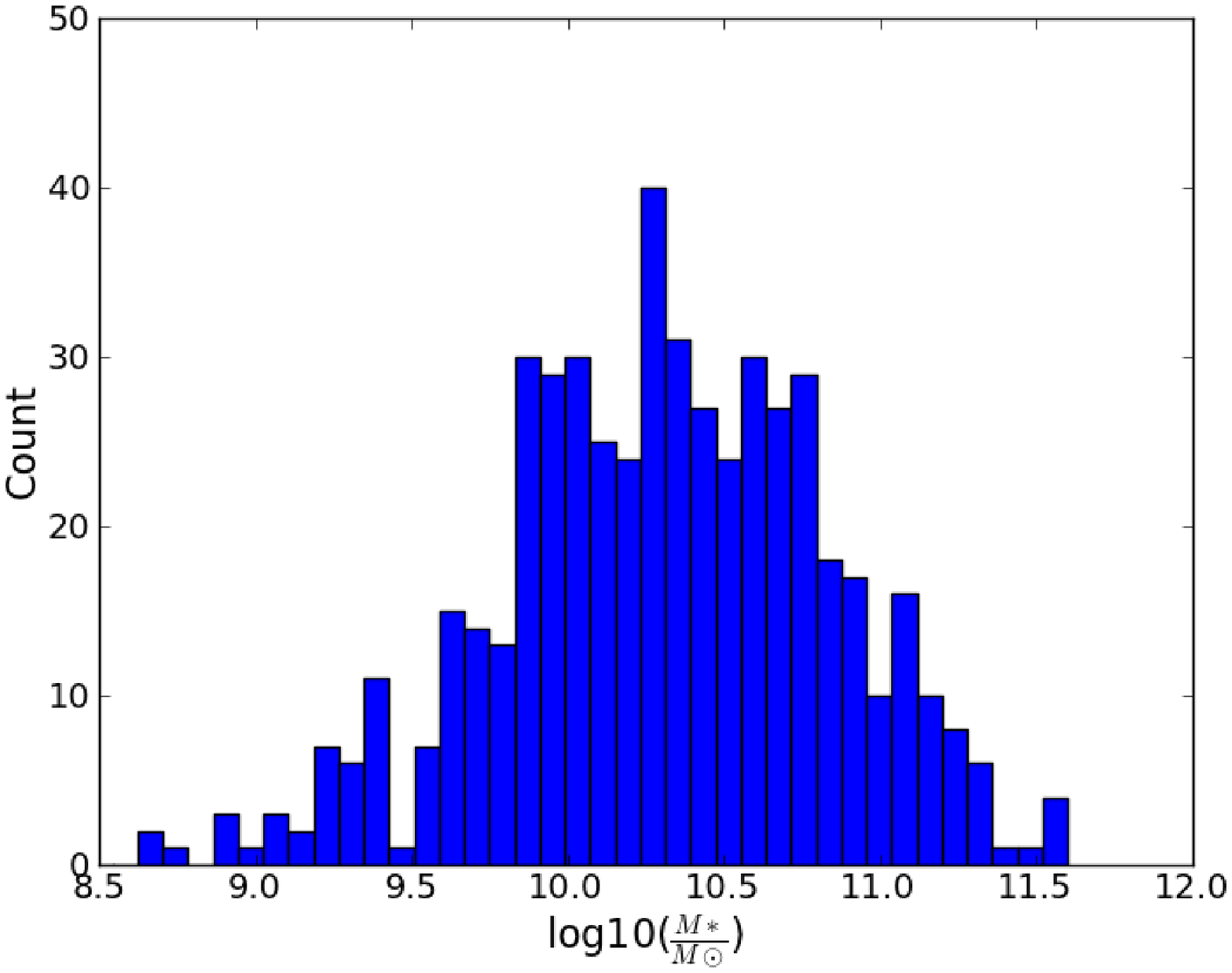}
  \caption{
Estimates of the stellar masses of the galaxies in sample 1 from modeling their spectral
energy distributions with MAGPHYS (Allen et al. in preparation).
}
\end{figure}

Fifth, there is the issue that ETGs with lower stellar masses, such
as dwarf elliptical galaxies, do have low values of the S\'ersic index
(Graham and Guzman 2003). Allen et al. (2015) have used MAGPHYS to estimate the stellar masses
of the H-GOODS galaxies using the photometry provided in Guo
et al. (2013). Figure 8 shows the distribution of stellar masses for the
HGOODS galaxies, which
shows that their low
values of the S\'ersic index cannot be attributed to low stellar
masses. 

With a few caveats, the most important of which is probably our
assumption of a universal initial mass function, our conclusion that most stars
form in LTGs but most of the stellar mass-density today is in ETGs seems secure.
A necessary implication of our conclusion 
is that there
must have been some process that transformed the morphologies of many galaxies
from the LTG class to the ETG class
after most of the stars in each galaxy had been formed.

However, before we consider the mechanisms that might cause this
major morphological transformation of the galaxy population,
we have to consider what our LTG/ETG classification means when we
apply it to galaxies at high redshift. At low redshift, an LTG is generally
a galaxy whose structure is dominated by a disk and an ETG is a galaxy
which has a prominent bulge or spheroid. A common assumption is
that this must also be true at high redshift, which is part of the reason
for the
conclusion from HST imaging studies that star-forming galaxies
at high redshift have structures dominated by disks
(e.g Wuyts et al. 2011; Bruce et al. 2012).
However, there is at least one way in which galaxy evolution might
have made this an unsafe assumption.
The bulges of low-redshift spiral galaxies
often have S\'ersic indices $<2.5$ (Balcells et al. 2003). Therefore, one
possibility we have to consider is that the high-redshift galaxies are
naked spheroids, which will subsequently accrete a disk and grow into
the spiral galaxies we see around us today. 

There are two main arguments againsts this idea.
The first is the circumstantial argument 
that most of the high-redshift galaxies do actually look
like they have disks.
In Figure 9 we have shown the CANDELS
images in the $H_{AB}$-band (1.6 $\mu$m) of the galaxies in sample 1
\footnote{We have chosen sample 1 rather than sample 2, despite the
fact the latter constitutes statistically
a larger fraction of the background radiation, because we can be more confident
that individual galaxies in sample 1 are emitting far-infrared
radiation since each was detected by the Herschel Space Observatory
at 160 $\mu$m.}. There are very few galaxies in this sample, even
ones with high values of the S\'ersic index, for which there is
not some sign of a disk on the HST image.

The second argument that our operationally-defined LTGs
are similar to the disk-dominated systems in the Universe today comes
from the results of
a different technique for investigating the morphologies of
high-redshift galaxies. A number of groups have inferred the morphological
composition of a high-redshift sample by modelling the 
distribution of ellipticities of the images of 
the galaxies (e.g. Ravindranath et al. 2006; Van der Wel et
al. 2014). Using a large sample of $\simeq$40,000 star-forming galaxies
from CANDELS, Van der Wel et al.
(2014) show that the distribution of ellipticities of star-forming CANDELS
galaxies with $1.5 < z < 2.0$ and stellar masses $>10^{10}\ M_{\odot}$
can be reproduced by a model in which $\simeq$75\% of the galaxies
are disks, with almost all the remaining objects having a spheroidal
structure. Both studies also found a significant population with highly
elongated structures which could not be explained by the predicted fraction
of highly edge-on disks, but in the study of the CANDELS sample these
highly elongated galaxies are found below a stellar mass of $10^{10}\ M_{\odot}$,
whereas most of the galaxies in our sample 1 have higher stellar masses
(Fig. 8).

Despite these arguments,
we can not rule
out the idea that high-redshift LTGs are physically a different type of
object to the disk-dominated galaxies in the local Universe,
especially as high-redshift galaxies are physically smaller
(Bruce et al. 2014) and have more turbulent
velocity fields (Genzel et al. 2014).
{\bf However, even if the LTGs at high redshift are physically distinct
from the disk-dominated objects that are LTGs at low redshift,
this does not eliminate the need to find a
mechanism for transforming LTGs into ETGs.} Suppose, for example,
the high-redshift LTGs are naked spheroids. There still needs to
be some mechanism to turn these into ETGs, since the obvious evolutionary
process, the growth of a
disk around the spheroid, would still produce a low-redshift 
galaxy with $n < 2.5$.

Before we turn to the mechanisms that might have caused this morphological shift,
it is worth pointing out that our result is a statistical result and does
not mean there can not be evolutionary mechanisms that either leave the morphologies
untouched or work in the other
direction.
For example, Graham et al. (2015) have argued that the compact passive galaxies
discovered at high redshift a decade ago (Daddi et al. 2005; Trujillo et al.
2006) are the ancestors of the bulges of low-redshift S0 or spiral galaxies,
and that the evolution from one into the other will occur by the
growth of a disk around the naked bulge - the same mechanism that
we considered above for the general galaxy 
population. 
It is possible that this morphological transformation occurs, but the following
argument implies that its scale is much less than the morphological
transformation from LTGs to ETGs implied by the calorimetry results.

Graham et al. have estimated that the descendant population
in this transformation,
the spirals and S0s in the Universe today with bulge structures very similar to the structures
of the compact passive galaxies at high redshift, have 
stellar masses of $\simeq10^{11}\ M_{\odot}$ and a volume density of
$3.5 \times 10^{-5}\ Mpc^{-3}\ dex^{-1}$, although this is strictly only
a lower limit. The
stellar mass-density in the descendant population  
is thus $\simeq3.5 \times 10^6\ M_{\odot}\ Mpc^{-3}$. However, the total stellar mass-density
in the Universe today (the integral of the distribution
in Fig. 2) is $1.9 \times 10^8\ M_{\odot}\ Mpc^{-3}$, approximately 50 times
larger. Thus this evolutionary channel may exist, but the morphological
change from LTGs to ETGs revealed by the calorimetry results involves a much larger fraction of the
current stellar mass-density.

We now consider the mechanisms that might have produced this major change, and
from now on we assume that the LTGs at high redshift are disk-dominated
galaxies.
Of the theoretical ideas for morphological transformation that
were discussed in \S 1, two are clearly consistent with the idea
that LTGs predate ETGs: 1) the idea that ETGs are formed by the merger
of two LTGs (Toomre 1977); (2) the idea that the spheroid in a galaxy is built
up by the rapid motion of star-forming clumps towards the centre of the
galaxy (Noguchi 1999; Bournaud et al. 2007; Genzel et al. 2011, 2014). 

The evidence from HST imaging that star-forming galaxies at all
redshifts are LTGs and passive galaxies are ETGs
(Bell et al. 2004a; Wuyts et al. 2011;
Bell et al. 2012;
Bruce et al. 2012; Buitrago et al. 2013; Szomoru et al. 2013;
Tasca et al. 2014) is evidence that the process responsible for
the morphological transformation is the same one that is
responsible for quenching the star formation in a galaxy
(Bell et al. 2012).
This evidence favours the second of the ideas above, since it
provides a clear explanation for both morphological
transformation and quenching.
Additional circumstantial evidence for
this conclusion is that
most ETGs in the
Universe today contain some evidence of a residual disk once 
you look closely enough (Krajnovic et al. 2013).

To explain our result, the
morphological transformation of an LTG must have occurred after most of the stars in the
galaxy had formed.
The average star-formation rate in the Universe dropped rapidly over the redshift range
$1 > z > 0$ (Hopkins and Beacom 2006; Eales et al. 2015), suggesting
that this redshift range may have been the one in which morphological transformation
was most active.
Two other results point in this direction. First, many studies have shown that
$\simeq$50\% of the stellar mass in `passive galaxies', which overlaps substantially
with ETGs, was formed since $z \simeq 1$ (Chen et al. 2003; Bell et al.
2004; Faber et al. 2006). Second, Tasca et al. (2014) have observed a gradual
decline in the optical emission from disks over this redshift range.
If this is the critical redshift range, the high-resolution wide-field surveys
with Euclid will be crucial for investigating the transformation
process.

\section{Conclusions}

In this paper, we have tried to quantify the importance of
morphological transformation in galaxy evolution.
We have used the operational
definition that an early-type galaxy (ETG) is one with a S\'ersic index $>2.5$ and a late-type
galaxy (LTG) is one with a S\'ersic index of $<$2.5. 
Using this definition, we have obtained the following two observational results:

\begin{itemize}

\item[1] We have used the results of the
Galaxy and Mass Assembly 
project
and the Herschel Astrophysical Terrahertz Large-Area Survey 
to show that in the Universe today 
51\% of the stellar mass-density is in ETGs
but 89\% of the rate of production of stellar mass-density
is
occurring in LTGs.

\item[2] The extragalactic background radiation contains 
all the energy generated by nuclear fusion in stars
since the Big Bang. We have resolved the background radiation into individual galaxies
using the deepest far-infrared survey with the Herschel Space Observatory and a deep
near-infrared/optical survey with the Hubble Space Telescope (HST). Using measurements
of the S\'ersic index of these galaxies derived from the HST images, we estimate that
$\simeq$83\% of the stellar mass-density formed over the history of the Universe occurred
in LTGs.

\end{itemize}

Our second result is subject to a number of caveats, the most importance of which is 
that our calorimetric method is based on the assumption of a universal initial
mass function. 
Subject to these caveats, the difference between the fraction of the 
stellar mass-density today that is in LTGs
and the fraction that was formed in these galaxies implies that there must have
been a major morphological transformation of LTGs into ETGs
after the formation of most of the stars. 

Our remaining conclusions are less certain. Since the star-formation density 
started to decline at $z \simeq 1$ and since there is evidence for the build-up
of stellar mass-density in passive galaxies since that redshift, it seems
likely that the morphological transformation from LTGs to ETGs occurred over
the redshift range $0<z<1$. 
If this conclusion is true,
a high-resolution investigation of the morphologies of galaxies
over this redshift range, such as will be possible with Euclid,
will be important for determining the process that is
responsible for the transformation.

The fact that HST imaging programmes show that star-forming galaxies
at all redshifts are dominated by disks, while passive galaxies have spheroidal
structures, implies there
is a single process responsible both for the morphological transformation and the
quenching of galaxies. This favours models such as those in which
the growth of a bulge in the centre
of a disk shuts down star formation by
stablising the disk against gravitational collapse.

\section*{Acknowledgments}

We thank the H-GOODS and CANDELS teams for producing such beautiful datasets and the
referee for an erudite and useful referee's report. 
LD, RI and SM acknowledge support from the European Research Council (ERC) 
in the form of the Advanced Investigator Program, 321302, COSMICISM.
The
Herschel-ATLAS is a project with Herschel, which is an ESA space observatory with science
instruments provided by European-led Principal Investigator consortia and with important
participation from NASA. The H-ATLAS website is http://www.h-atlas.org/. GAMA is a joint
European-Australasian project based around a spectroscopic campaign using the Anglo-
Australian Telescope. The GAMA input catalogue is based on data taken from the Sloan Digital
Sky Survey and the UKIRT Infrared Deep Sky Survey. Complementary imaging of the GAMA
regions is being obtained by a number of independent survey programs, including GALEX, MIS,
VST KIDS, VISTA VIKING, WISE, Herschel-ATLAS, GMRT and ASKAP, providing UV to
radio coverage. GAMA is funded by the STFC (UK), the ARC (Australia), the AAO, and the
participating institutions. The GAMA website is http://www.gama-survey.org/. HerMES is a
Herschel Key Programme utilizing Guaranteed Time from the SPIRE instrument team, ESAC
scientists and a mission scientist.

\begin{figure*}
\includegraphics[width=140mm]{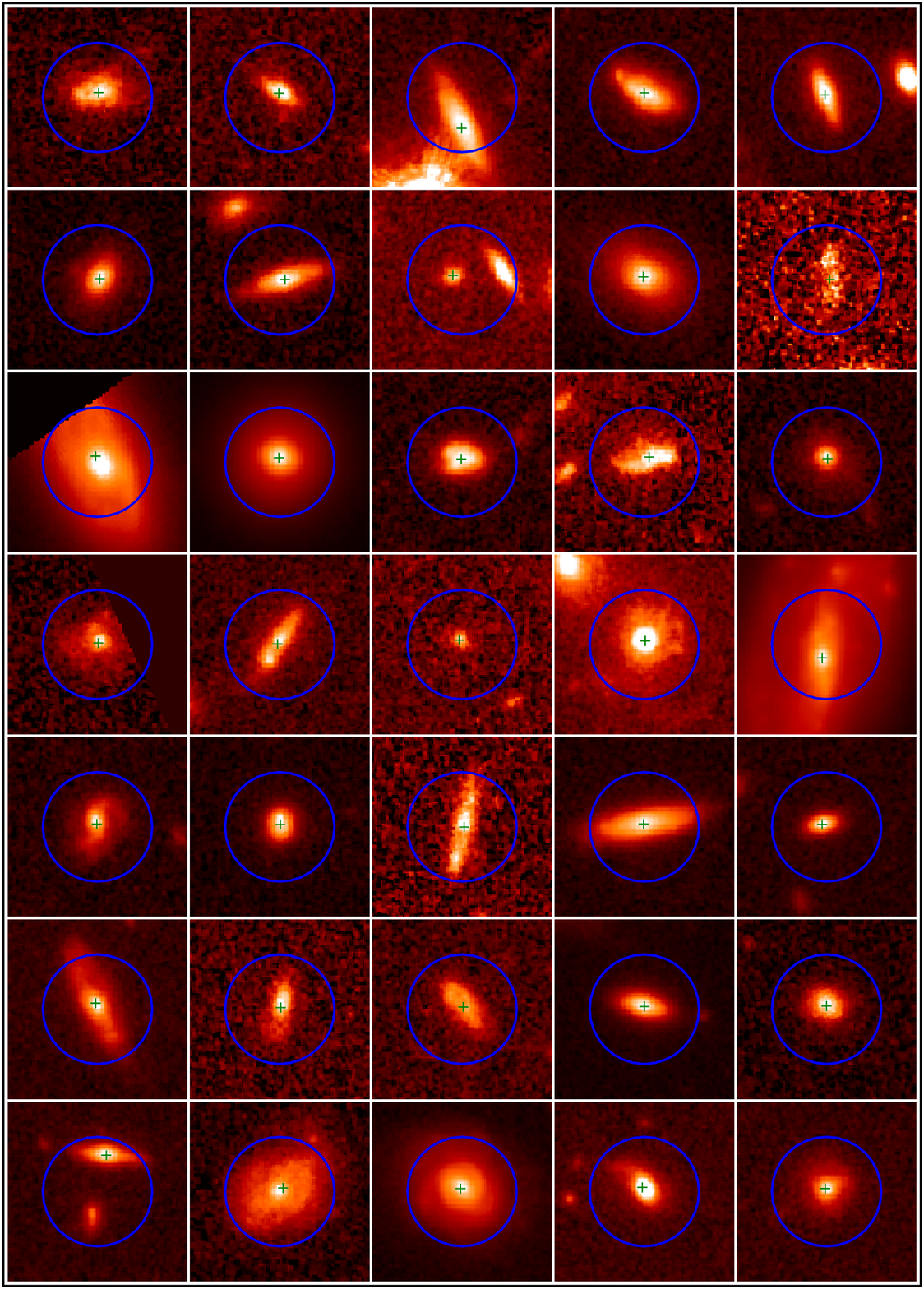}
  \caption{The H-band images (1.6 $\mu$m) of each of the Herschel 
sources in sample 1. The
images are centred on the Herschel positions and have a size of 
$5\times5\ arcsec^2$. The cross shows
the near-infrared counterpart to the Herschel source that we 
have identified using our probabalistic
analysis. The circle has a radius of 5 arcsec (the remaining sources are shown in
the on-line version of the paper).
}
\end{figure*}

\begin{figure*}
\includegraphics[width=140mm, trim=-30mm 0mm 0mm 0mm]{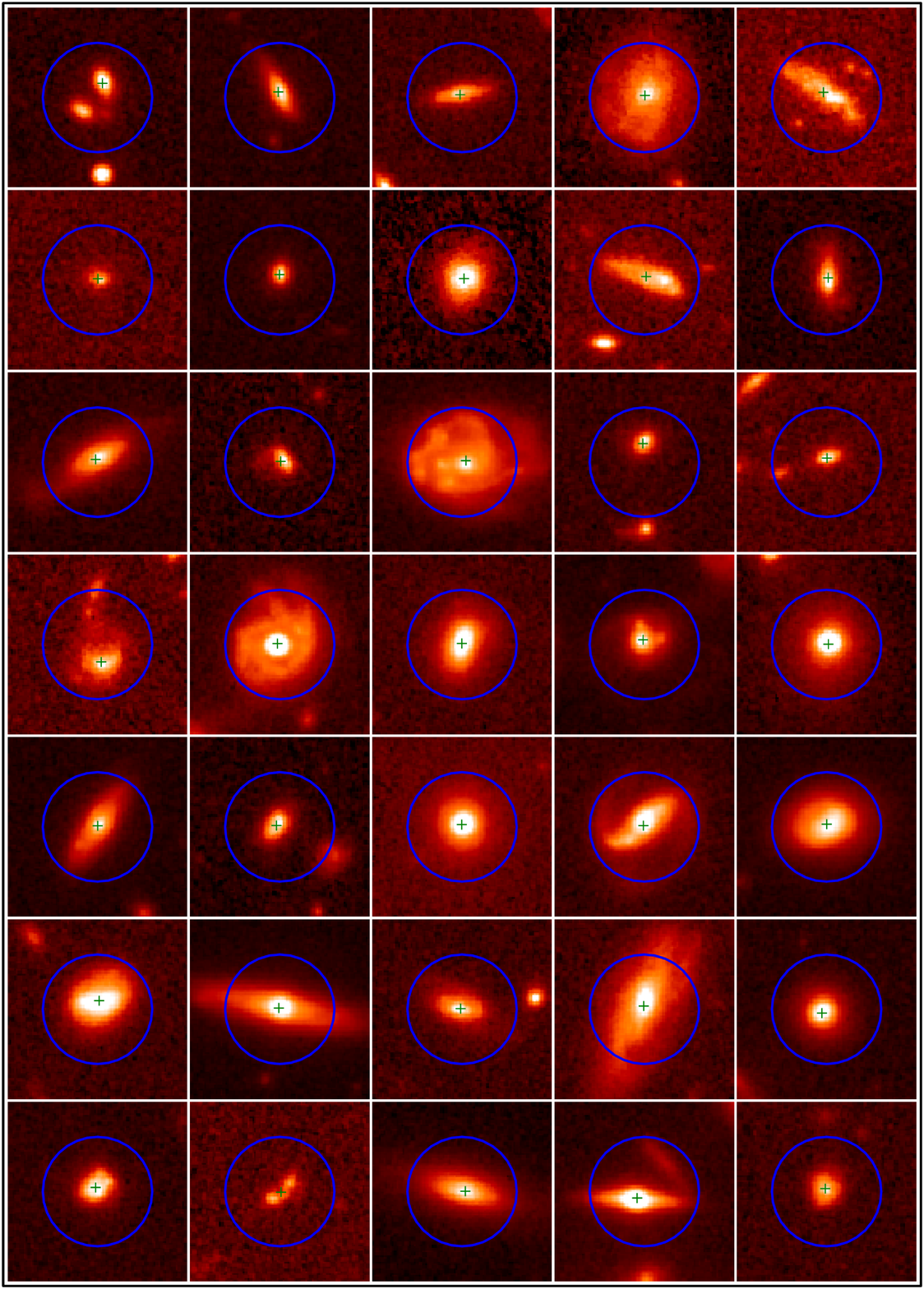}
  \contcaption{}
\end{figure*}

\begin{figure*}
\includegraphics[width=140mm, trim=-30mm 0mm 0mm 0mm]{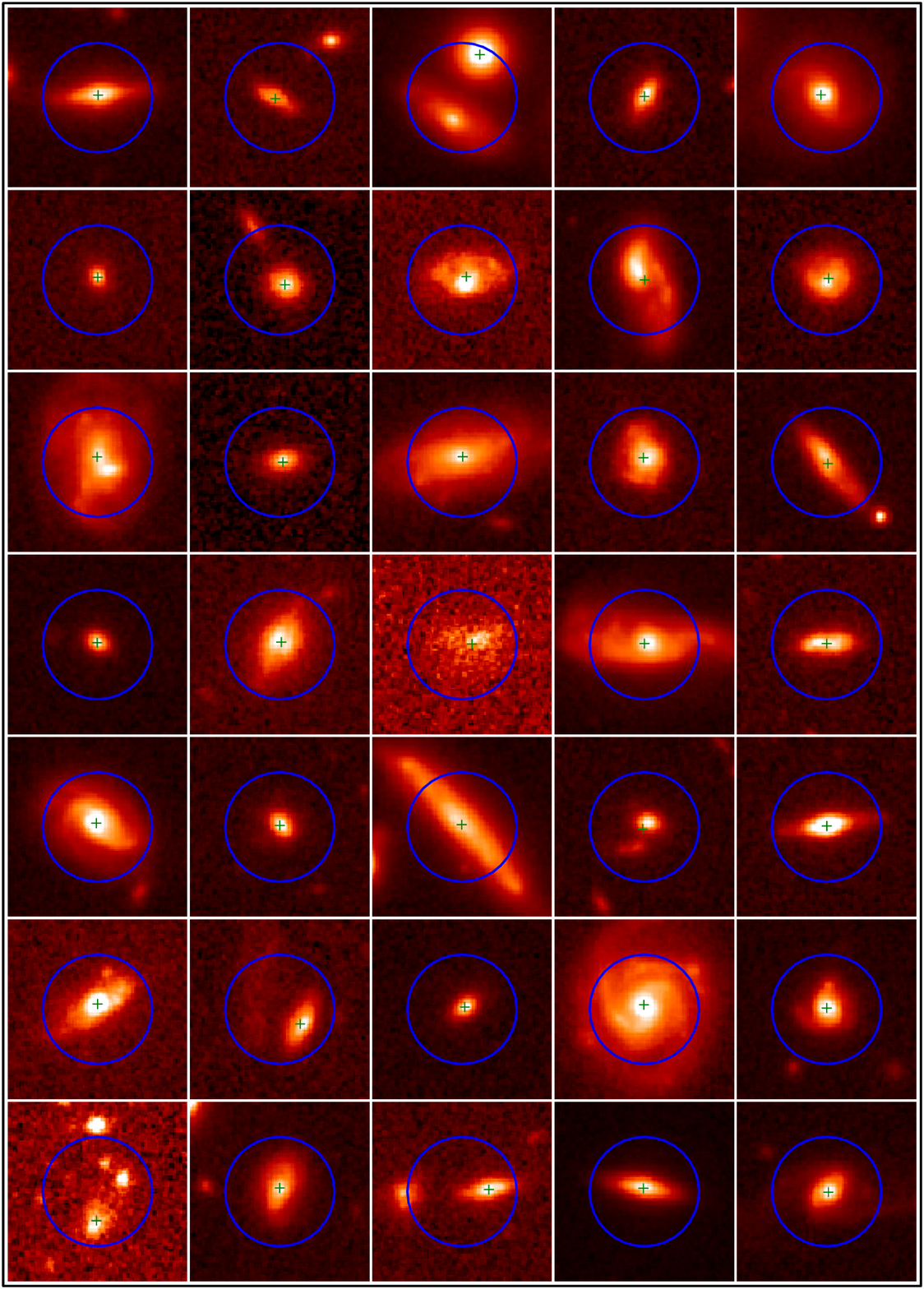}
  \contcaption{}
\end{figure*}

\begin{figure*}
\includegraphics[width=140mm, trim=-30mm 0mm 0mm 0mm]{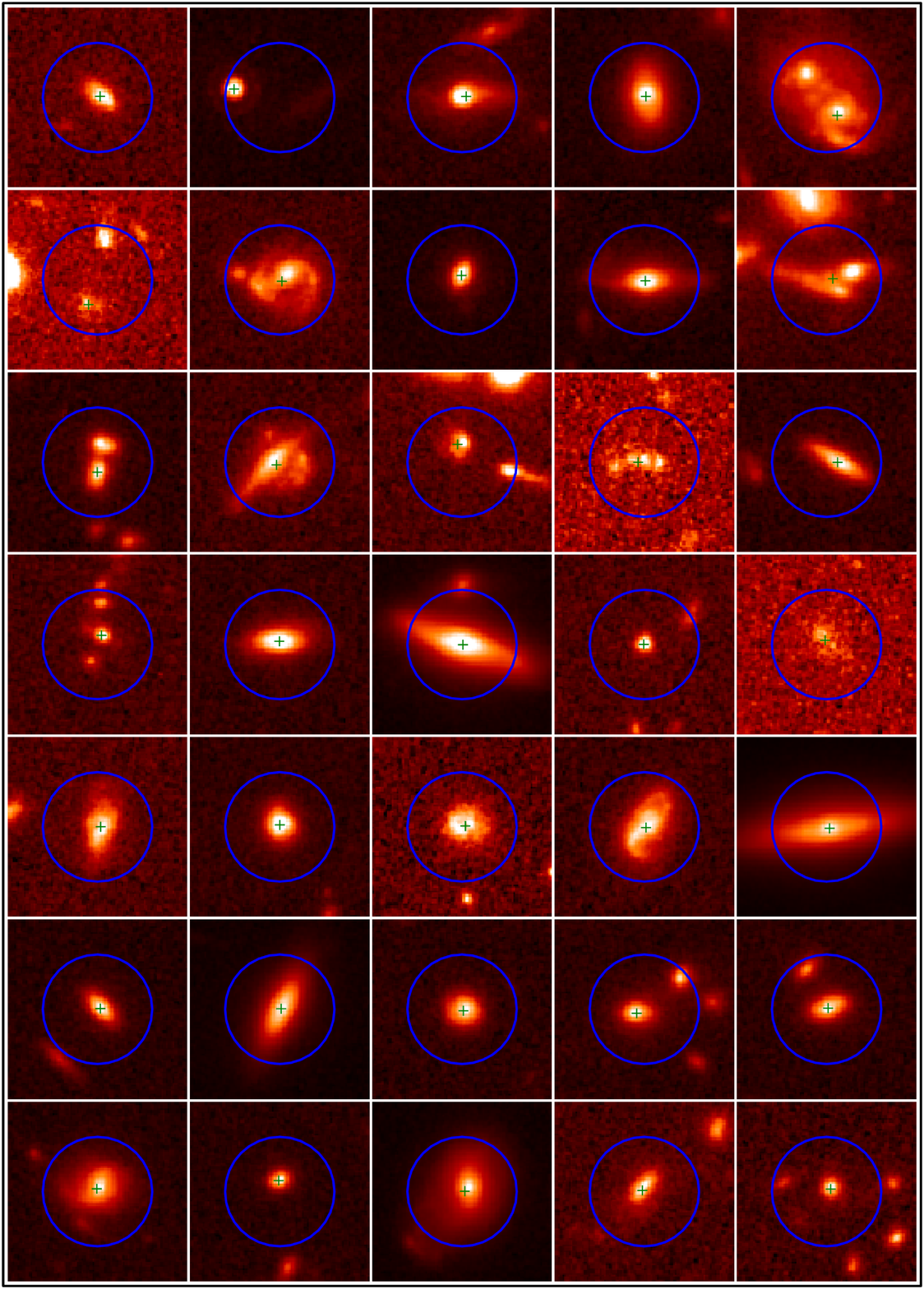}
  \contcaption{}
\end{figure*}

\begin{figure*}
\includegraphics[width=140mm, trim=-30mm 0mm 0mm 0mm]{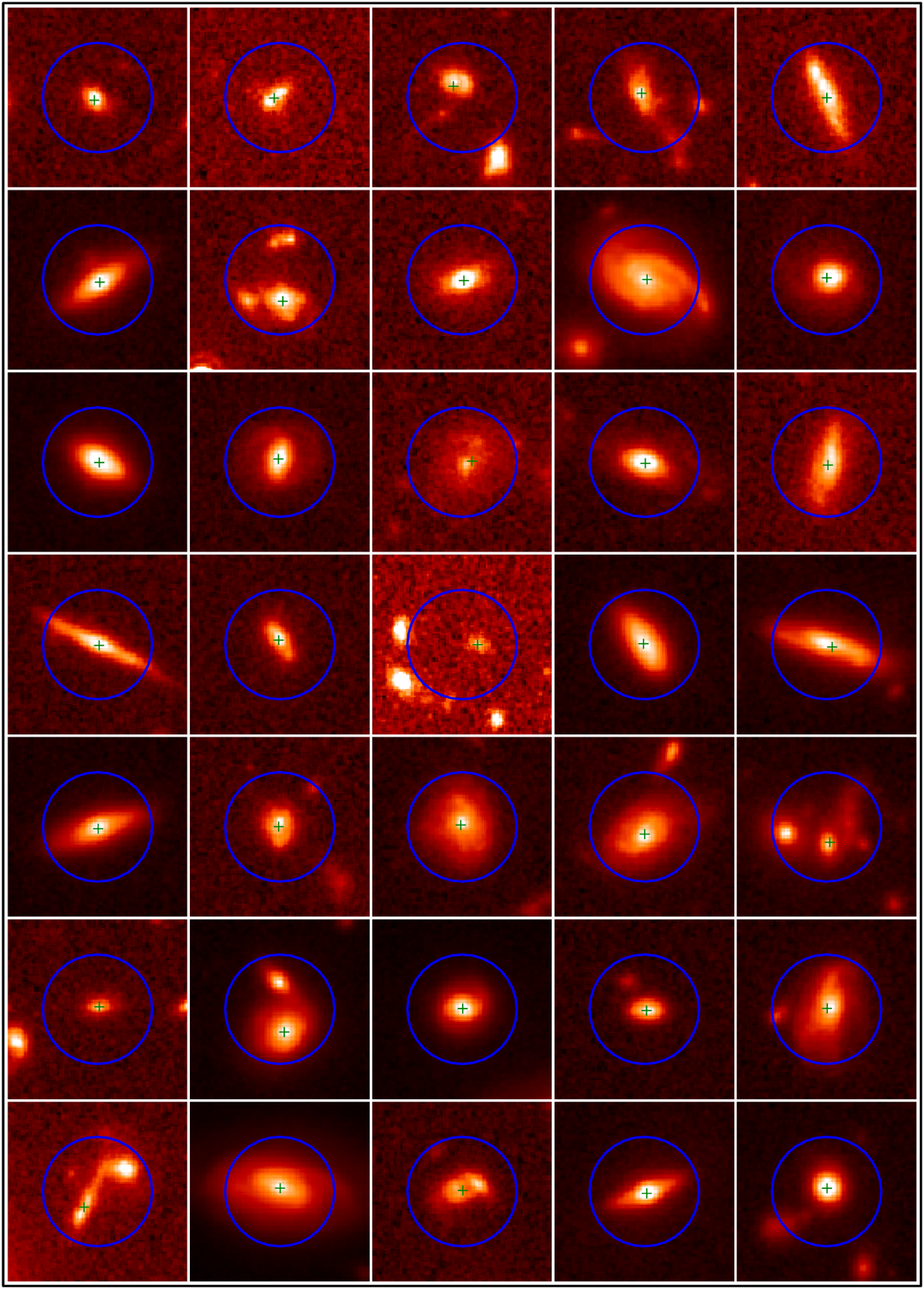}
  \contcaption{}
\end{figure*}

\begin{figure*}
\includegraphics[width=140mm, trim=-30mm 0mm 0mm 0mm]{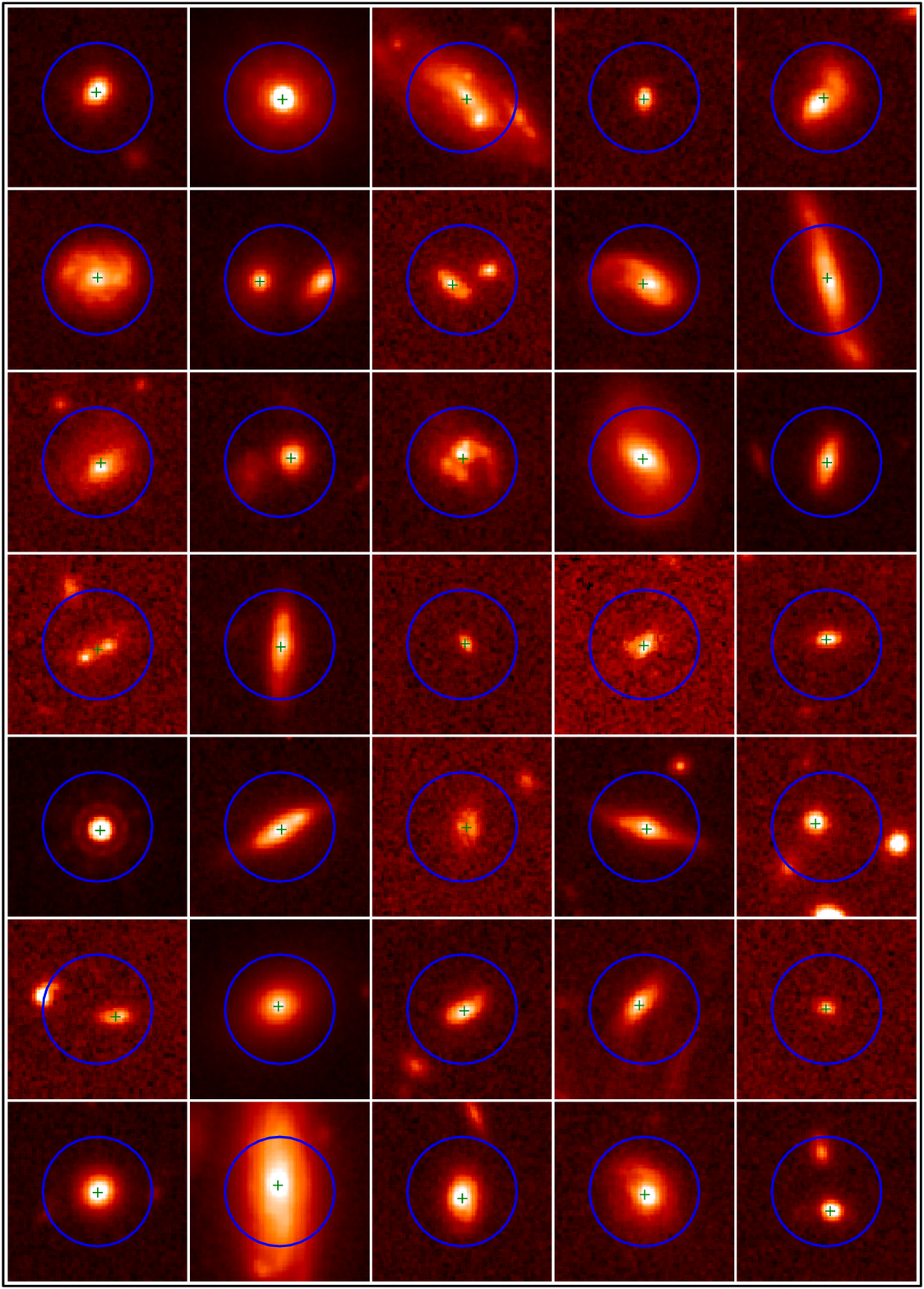}
  \contcaption{}
\end{figure*}

\begin{figure*}
\includegraphics[width=140mm, trim=-30mm 0mm 0mm 0mm]{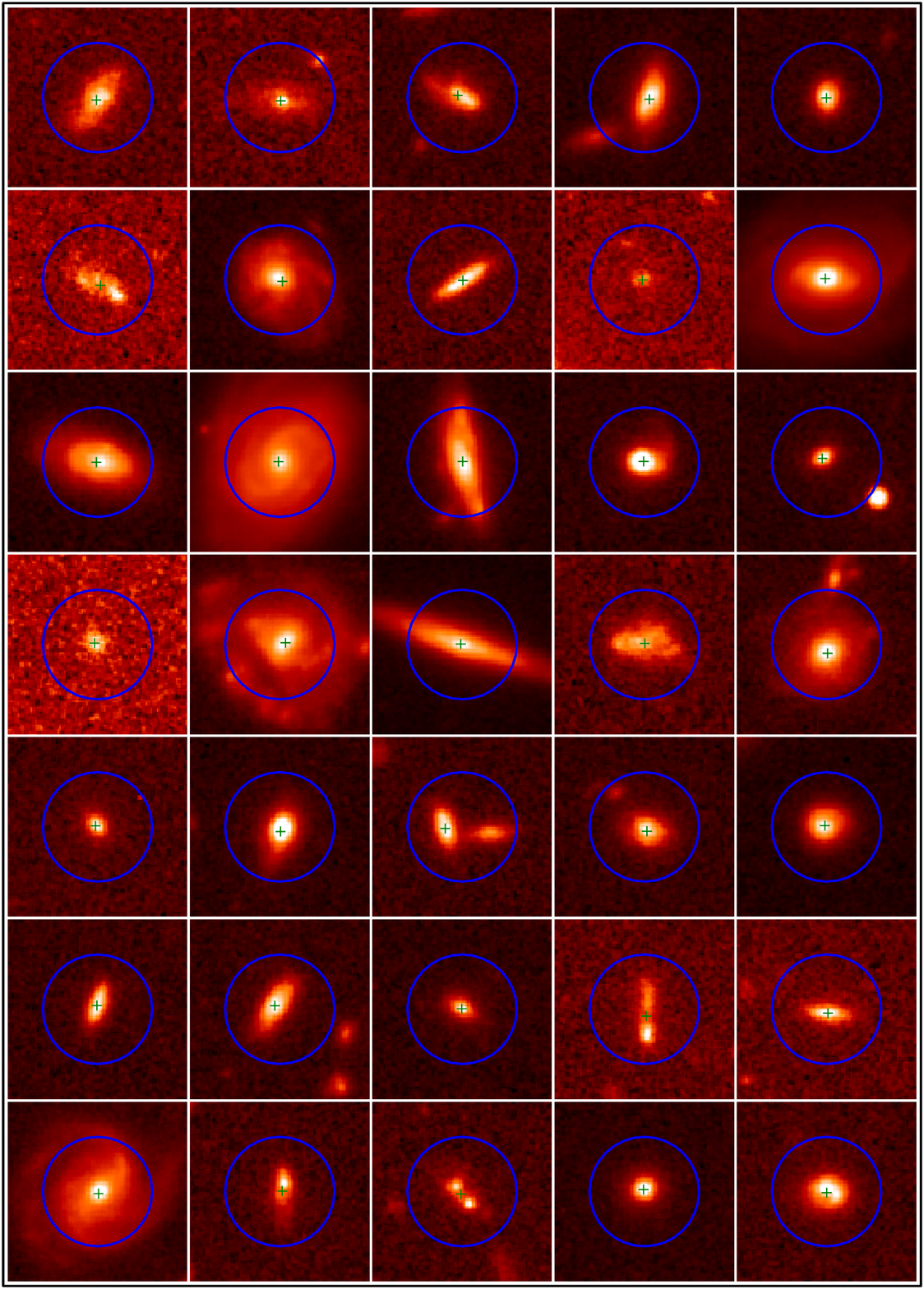}
  \contcaption{}
\end{figure*}

\begin{figure*}
\includegraphics[width=140mm, trim=-30mm 0mm 0mm 0mm]{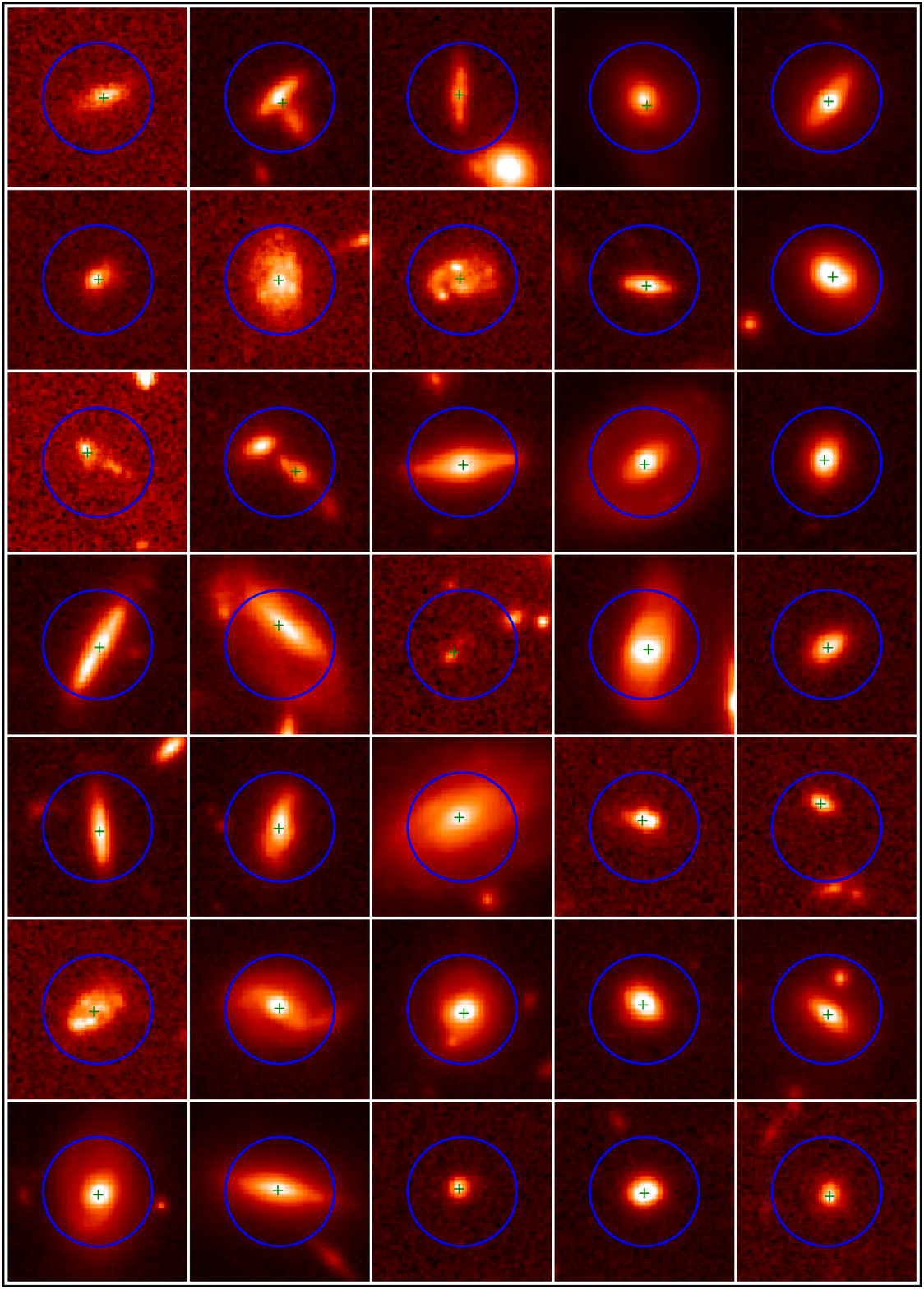}
  \contcaption{}
\end{figure*}

\begin{figure*}
\includegraphics[width=140mm, trim=-30mm 0mm 0mm 0mm]{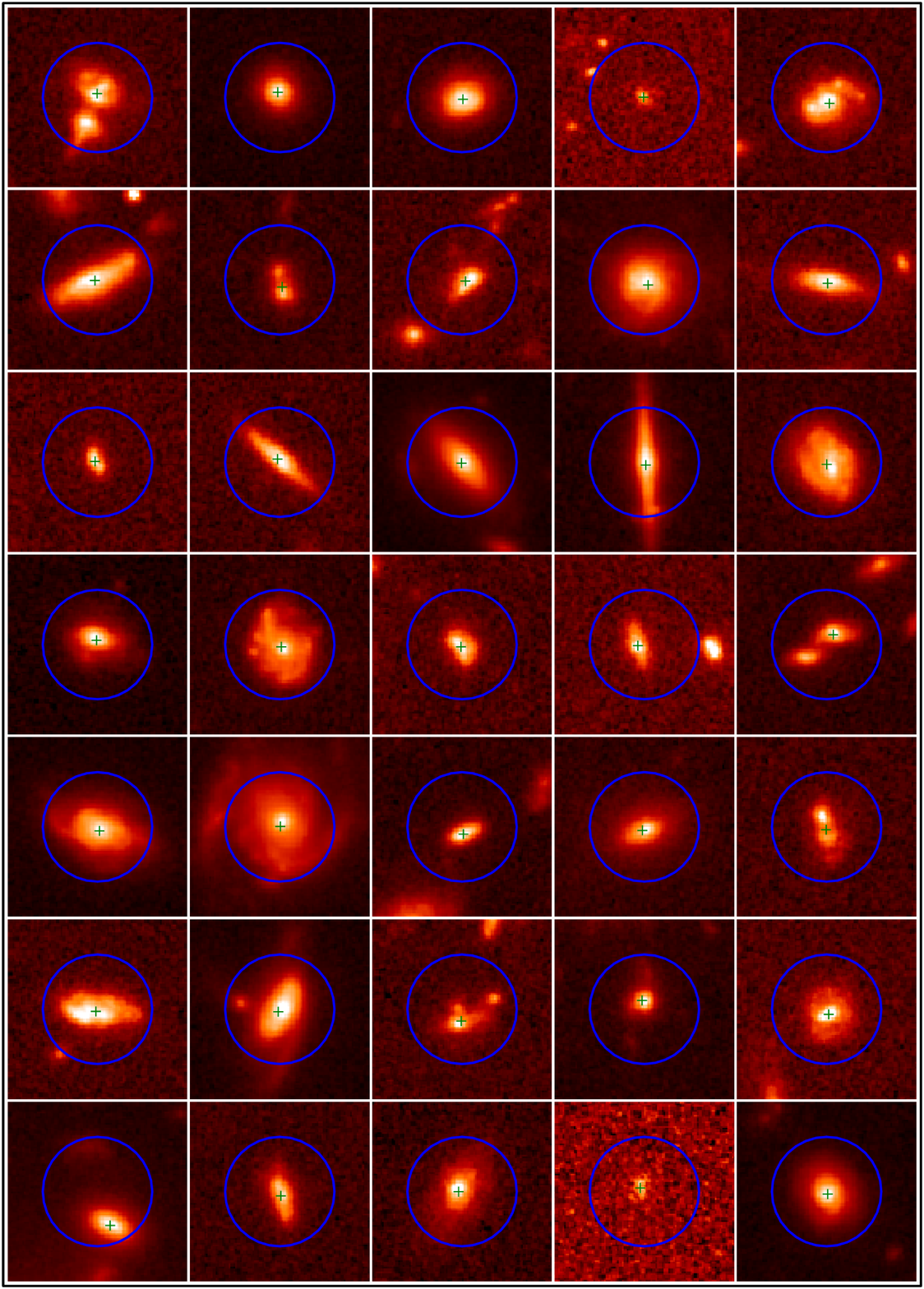}
  \contcaption{}
\end{figure*}

\begin{figure*}
\includegraphics[width=140mm, trim=-30mm 0mm 0mm 0mm]{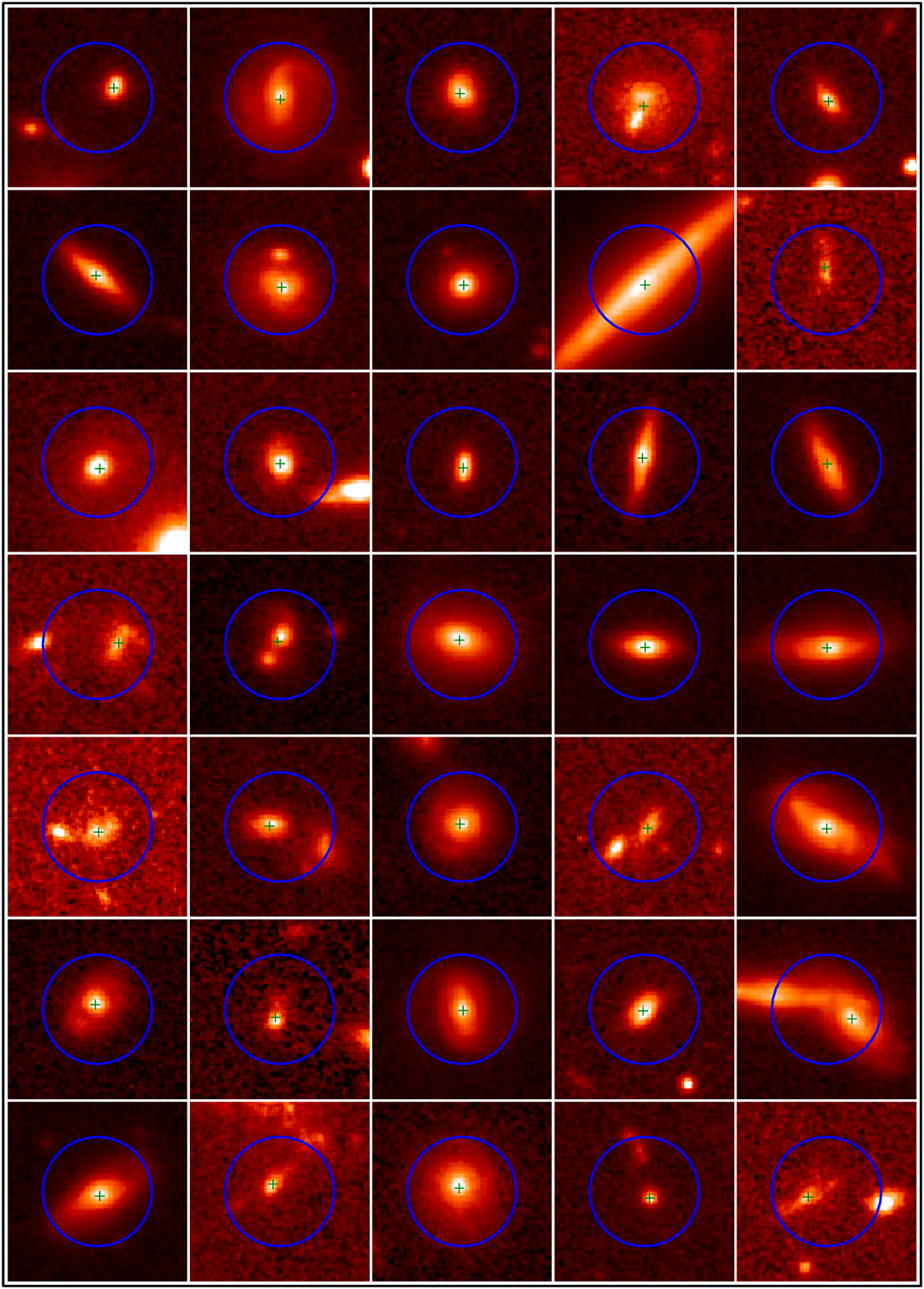}
  \contcaption{}
\end{figure*}

\begin{figure*}
\includegraphics[width=140mm, trim=-30mm 0mm 0mm 0mm]{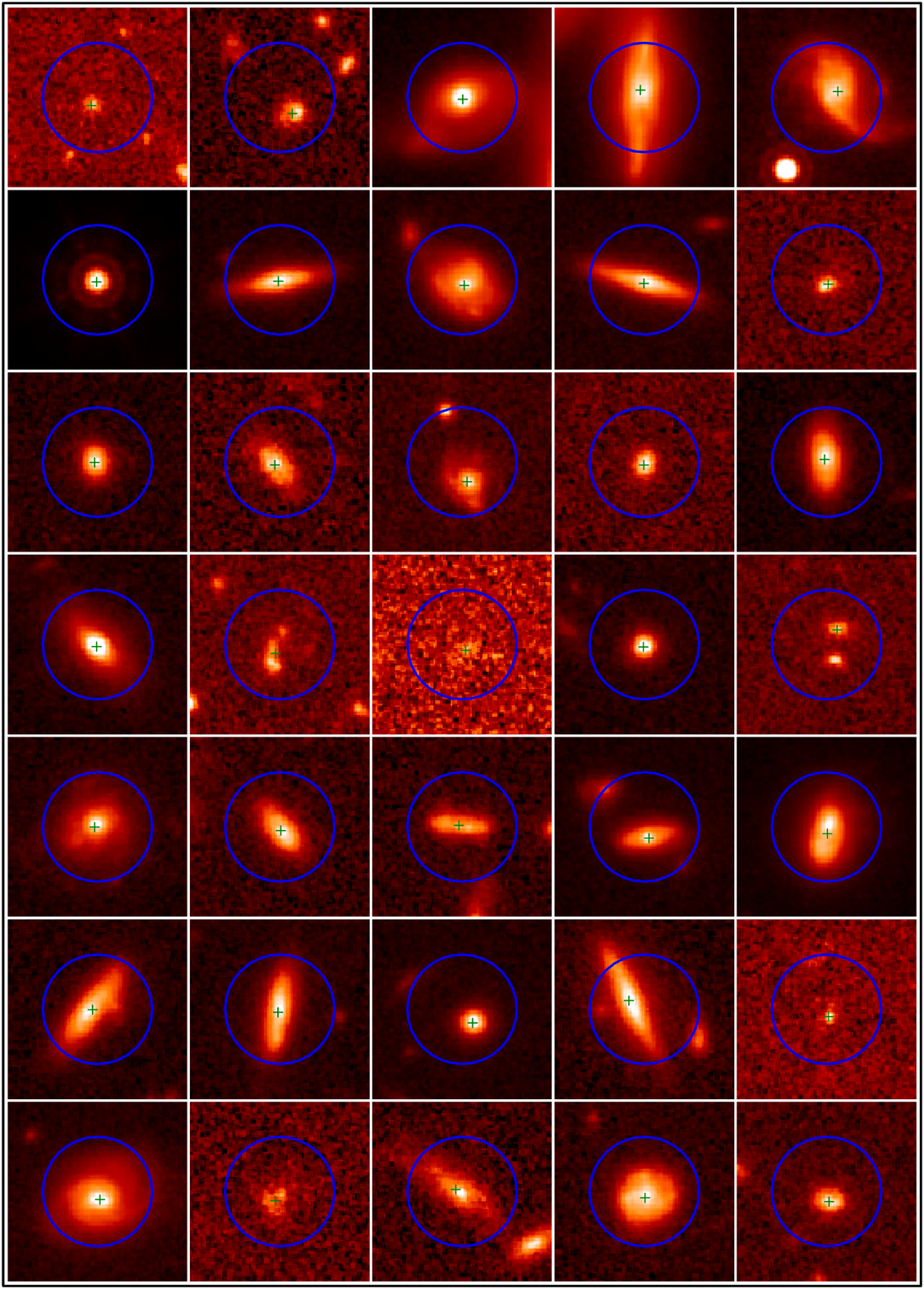}
  \contcaption{}
\end{figure*}

\begin{figure*}
\includegraphics[width=140mm, trim=-30mm 0mm 0mm 0mm]{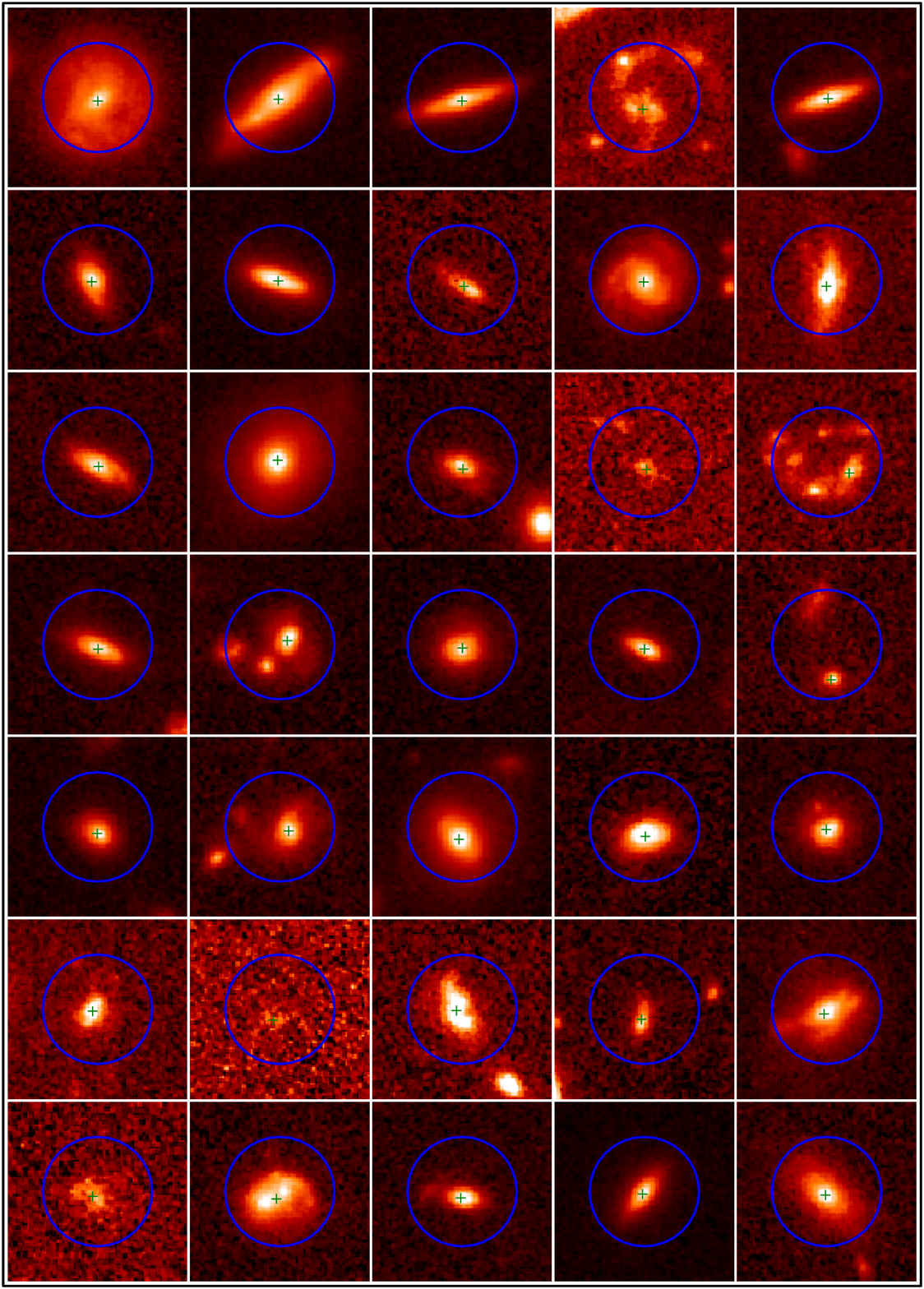}
  \contcaption{}
\end{figure*}


\begin{thebibliography}{99}

\bibitem[\protect\citeauthoryear{Abazajian et al.}{2009}]{ab1} Abazajian, K. et al.
2009, ApJS, 182, 543
\bibitem[\protect\citeauthoryear{Adami et al.}{2005}]{ad} Adami, C. et al.
2005, A.\&A. 443, 805
\bibitem[\protect\citeauthoryear{Balcells et al.}{2003}]{bal} Balcells, M.,
Graham, A.W., Dominguez-Palmero, L. \& Peletier, R.F. 2003,
ApJ, 582, L79
\bibitem[\protect\citeauthoryear{Baldry}{2008}]{bald1} Baldry, I.K. 2008,
A\&G, 49, 25
\bibitem[\protect\citeauthoryear{Baldry}{2012}]{bald} Baldry, I.K. et al.
2012, MNRAS, 421, 621
\bibitem[\protect\citeauthoryear{Bell}{2004a}]{bella} Bell, E.F. et al.
2004a, ApJ, 600, L11
\bibitem[\protect\citeauthoryear{Bell}{2004b}]{bellb} Bell, E.F. et al.
2004b, ApJ, 608, 752
\bibitem[\protect\citeauthoryear{Bell}{2012}]{bellc} Bell, E.F. et al.
2012, ApJ, 753, 167
\bibitem[\protect\citeauthoryear{Bendo}{2015}]{bendo} Bendo, G. et al.
2015, MNRAS, 448, 135
\bibitem[\protect\citeauthoryear{Barden}{2005}]{bard} Barden, M. et al.
2005, ApJ, 635, 959
\bibitem[\protect\citeauthoryear{Bournaud}{2007}]{bourn} Bournaud, F., Elmegreen,
B.G. and Elmegreen, D.M. 2007, ApJ, 670, 237
\bibitem[\protect\citeauthoryear{Bourne}{2015}]{bourne} Bourne, N. et al.
2015, in preparation
\bibitem[\protect\citeauthoryear{Bruce}{2012}]{bruce} Bruce, V.A. et al.
2012, MNRAS, 427, 1666
\bibitem[\protect\citeauthoryear{Bruce}{2014}]{bruce2} Bruce, V.A. et al.
2014, MNRAS, 444, 1660 
\bibitem[\protect\citeauthoryear{Buitrago}{2013}]{buit} Buitrago,
F., Trujillo, I. Conselice, C.J. \& Haussler, B. 2013, MNRAS, 428, 1460
\bibitem[\protect\citeauthoryear{Burgarella}{2013}]{burg} Burgarella, D. et al.
2013, A.\&A., 554, 70
\bibitem[\protect\citeauthoryear{Cambresy}{2001}]{cam} Cambresy, L., Reach, W.T.,
Beichman, C.A. \& Jarrett, T.H. 
2001, ApJ, 555, 563
\bibitem[\protect\citeauthoryear{Chen}{2003}]{chen} Chen, H.-W. et al. 2003,
ApJ, 586, 745
\bibitem[\protect\citeauthoryear{Combes}{2014}]{combes} Combes, F. 2014,
{\it Structure and Dynamics of Disk Galaxies}, eds M.S. Seigar and P. Treuthardt,
ASP Conference Series, 480, 211
\bibitem[\protect\citeauthoryear{Cortese}{2012}]{luca} Cortese, L. 2012,
A\&A, 543, 132
\bibitem[\protect\citeauthoryear{Da Cunha}{2008}]{dac} Da Cunha, E., Charlot,
S. \& Elbaz, D. 2008, MNRAS, 388, 1595
\bibitem[\protect\citeauthoryear{Daddi}{2005}]{dad} Daddi, E. et al. 2005, ApJ,
626, 680
\bibitem[\protect\citeauthoryear{de Vaucouleur}{1948}]{dev} de Vaucouleurs, G. 1948,
Ann d'Ap, 11, 247 
\bibitem[\protect\citeauthoryear{Dekel}{2014}]{dek} Dekel, A. \&
Burkert, A. 2014, MNRAS, 438, 1870
\bibitem[\protect\citeauthoryear{Dole}{2006}]{dole} Dole, H. et al.
2006, A\&A, 451, 417
\bibitem[\protect\citeauthoryear{Driver}{2008}]{drive2} Driver, S.P., Popescu,
C.C., Tuffs, R.J., Graham, A., Liske, J. \& Baldry,
I. 2008, ApJL, 678, L101
\bibitem[\protect\citeauthoryear{Driver}{2011}]{drive} Driver, S.P. et al.
2011, MNRAS, 413, 971
\bibitem[\protect\citeauthoryear{Dwek}{2013}]{dwek} Dwek, E. \& Krennrich, F.
2013, Astroparticle Physics, 43, 112
\bibitem[\protect\citeauthoryear{Dye}{2009}]{dye} Dye, S. et al. 2009,
ApJ, 703, 285
\bibitem[\protect\citeauthoryear{Eales}{2010}]{eales} Eales, S.A. et al.
2010, PASP, 891, 499 
\bibitem[\protect\citeauthoryear{Eales}{2015}]{eales} Eales, S.A. et al.
2015, MNRAS, in preparation
\bibitem[\protect\citeauthoryear{Emsellem}{2011}]{ems} Emsellem, E. et al.
2011, MNRAS, 414, 888
\bibitem[\protect\citeauthoryear{Elbaz}{2011}]{elbaz} Elbaz, D. et al. 2011,
A\&A, 533, 119
\bibitem[\protect\citeauthoryear{Faber}{2007}]{faber} Faber, S.M. et al.
2007, ApJ, 665, 265
\bibitem[\protect\citeauthoryear{Fixsen}{1998}]{fixsen} Fixsen, D.J. et al. 1998,
ApJ, 508, 128
\bibitem[\protect\citeauthoryear{Genzel}{2011}]{genzel1} Genzel, R. et al.
2011, ApJ, 733, 30
\bibitem[\protect\citeauthoryear{Genzel}{2014}]{genzel2} Genzel, R. et al.
2014, ApJ, 785, 75
\bibitem[\protect\citeauthoryear{Graham}{2003}]{graham2} Graham, A.W. \& Guzman, R.
2003, AJ, 125, 2936
\bibitem[\protect\citeauthoryear{Graham}{2005}]{graham3} Graham, A.W. and Driver,
S.P. 2005, PASA, 22, 118
\bibitem[\protect\citeauthoryear{Graham}{2008}]{graham} Graham, A.W.
and Worley, C.C. 2008, MNRAS, 388, 1708
\bibitem[\protect\citeauthoryear{Graham}{2013}]{graham2} Graham, A.W. 2013,
in {\it Planets, Stars and Stellar Systems}, Volume 6, p91-140,
eds T.D. Oswalt and W.C. Keel, Springer Publishing
(arXiV: 1108.0997)
\bibitem[\protect\citeauthoryear{Graham}{2015}]{graham3} Graham, A.W., 
Dullo, B.T. \& Savorgnan, G.A.D., MNRAS, in press (arXiv: 1502.07024)
\bibitem[\protect\citeauthoryear{Guo}{2013}]{guo} Guo, Y. et al.
2013, ApJ Suppl. 207, 24
\bibitem[\protect\citeauthoryear{Hopkins}{2006}]{hop} Hopkins, A.
\& Beacom, J.F. 2006, ApJ, 651, 142
\bibitem[\protect\citeauthoryear{Hubble}{1926}]{hub1} Hubble, E.P. 1926, ApJ, 64,
321
\bibitem[\protect\citeauthoryear{Hubble}{1927}]{hub2} Hubble, E.P. 1927, The Observatory,
50, 276
\bibitem[\protect\citeauthoryear{Ibar et al.}{2010}]{ib1} Ibar, E. et al. 2010, MNRAS, 409, 38
\bibitem[\protect\citeauthoryear{Ilbert et al.}{2013}]{ilb} Ilbert, O. et al.
2013, A.\&A. 556, 55
\bibitem[\protect\citeauthoryear{Kelvin}{2012}]{kel} Kelvin, L.S. et al.
2012, MNRAS, 421, 1007
\bibitem[\protect\citeauthoryear{Kelvin}{2014a}]{kel14a} Kelvin, L.S. et al.
2014a, MNRAS, 439, 1245
\bibitem[\protect\citeauthoryear{Kelvin}{2014b}]{kel14b} Kelvin, L.S. et al.
2014b, MNRAS, 444, 1647
\bibitem[\protect\citeauthoryear{Kennicutt}{1998}]{ke1} Kennicutt, R.C. 1998,
ARAA, 36, 189
\bibitem[\protect\citeauthoryear{Kennicutt and Evans}{2012}]{ke} Kennicutt,
R.C. \& Evans, N.J. 2012, ARAA, 50, 531
\bibitem[\protect\citeauthoryear{Koekemoer}{2011}]{kok} Koekemoer et al. 2011,
ApJ Suppl. 197, 36
\bibitem[\protect\citeauthoryear{Krajnovic}{2013}]{kraj} Krajnovic, D. et al.
2013, MNRAS, 432, 1768
\bibitem[\protect\citeauthoryear{Laurikainen}{2010}]{laur} Laurikainen, E., Salo, H.,
Buta, R., Knapen, J.H. \& Comer\'on, S. 2010, MNRAS, 405, 1089
\bibitem[\protect\citeauthoryear{Lee}{2013}]{lee} Lee, B. et al. 2013, ApJ, 774, 47.
\bibitem[\protect\citeauthoryear{Leitner}{2011}]{lei} Leitner, S.N. \& Kravtsov, A.V.
2011, ApJ, 734, 48
\bibitem[\protect\citeauthoryear{Leiton}{2015}]{lei} Leiton, R. et al. 2015, A.\&A. in press
(arXiv: 1503.05779)
\bibitem[\protect\citeauthoryear{Levenson}{2007}]{leven} Levenson, L.R., Wright,
E.L., Johnson, B.D. 2007, ApJ, 666, 34
\bibitem[\protect\citeauthoryear{Li-Ting}{2014}]{li} Li-Ting, H.
et al. 2014, arXiv: 1409.7119v1
\bibitem[\protect\citeauthoryear{Lilly}{1987}]{lilly} Lilly, S.J. \& Cowie,
L.L. 1987, {\it Infrared astronomy with arrays, Proceedings
of the Workshop on Ground-based Astronomical Observations with Infrared
Array Detectors, held at University of Hawaii, Hilo, March 24-27, 1987,} edited
by C.G. Wynn-Williams and E.E. Becklin, 473
\bibitem[\protect\citeauthoryear{Magnelli}{2013}]{mag} Magnelli, B. et al.
2013, A\&A, 553, 132
\bibitem[\protect\citeauthoryear{Martig}{2009}]{mart} Martig, M.,
Bournaud, F., Teyssier, R. \& Dekel, A. 2009, ApJ, 707, 250
\bibitem[\protect\citeauthoryear{Maraston}{2006}]{mar} Maraston, C. 2006,
MNRAS, 362, 799
\bibitem[\protect\citeauthoryear{Marsden}{2009}]{mars} Marsden, G. et
al. 2009, ApJ, 707, 1729
\bibitem[\protect\citeauthoryear{Martin et al.}{2005}]{ma1} Martin, D. et al. 2005, ApJ, 619, L1
\bibitem[\protect\citeauthoryear{Muzzin et al.}{2013}]{muz} Muzzin, A. et al. 2013, ApJ, 777, 18
\bibitem[\protect\citeauthoryear{Nguyen}{2010}]{nug} Nguyen, H.T. et al.
2010, A\&A, 518, L5
\bibitem[\protect\citeauthoryear{Noguchi}{1999}]{nog} Noguchi, M. 1999, ApJ, 514, 77
\bibitem[\protect\citeauthoryear{Oliver}{2012}]{oliver} Oliver, S. et al.
2012, MNRAS, 424, 1635
\bibitem[\protect\citeauthoryear{Pagel}{1997}]{pag} Pagel, B. 1997,
{\it Nucleosynthesis and Chemical Evolution of Galaxies} (Cambridge)
\bibitem[\protect\citeauthoryear{Pascale et al.}{2011}]{p1} Pascale, E. et al. 2011, MNRAS,
415, 911
\bibitem[\protect\citeauthoryear{Pilbratt}{2010}]{pil} Pilbratt, G. et al.
2010, A\&A, 518, L1
\bibitem[\protect\citeauthoryear{Peng}{2010}]{peng} Peng, Y.-J. et al.
2010, ApJ, 721, 193
\bibitem[\protect\citeauthoryear{Planck}{2013}]{pl} Planck Collaboration XVI 2014,
A \& A, 571, 16
A\&A submitted, arXiv: 1303.5706
\bibitem[\protect\citeauthoryear{Ravindranath et al.}{2006}]{r2} Ravindranath et al.
2006, ApJ, 652, 963
\bibitem[\protect\citeauthoryear{Rigby et al.}{2011}]{r1} Rigby, E.E. 
et al. 2011, MNRAS, 415, 2336
\bibitem[\protect\citeauthoryear{Sersic}{1963}]{ser} S\'ersic, J.L. 1963,
Boletin de la Asociacion Argentina de Astronomia, 6, 41
\bibitem[\protect\citeauthoryear{Shen}{2003}]{shen} Shen, S., Mo, H.J., White,
S.D.M., Blanton, M.R., Kauffmann, G., Voges, R., Brinkmann, J. \&
Csabai, I. 2003, MNRAS, 343, 978
\bibitem[\protect\citeauthoryear{Szomoru et al.}{2011}]{szo} Szomoru, D., Franx, M.,
Bouwens, R.J., Van Dokkum, P.G., Labb\'e, I., Illingworth, G.D. \& Trenti,
M. 2011, ApJ, 735, L22 
\bibitem[\protect\citeauthoryear{Smith et al.}{2011}]{s1} Smith, D.J.B. et al. 2011, MNRAS,
416, 857
\bibitem[\protect\citeauthoryear{Smith}{2012}]{smitha} Smith, M.W.L. et al. 2012,
ApJ, 748, 123
\bibitem[\protect\citeauthoryear{Smith}{2012}]{smithb} Smith, D.J.B. et
al. 2012b, MNRAS, 427, 703 
\bibitem[\protect\citeauthoryear{Tasca}{2014}]{tas} Tasca, L.A.M. et al. 2014,
A\&A, 564, 12
\bibitem[\protect\citeauthoryear{Takeuchi}{2005}]{tak} Takeuchi, T.T., Buat,
V. \& Burgarella, D. 2005, A.\&A., 440, L17
\bibitem[\protect\citeauthoryear{Taylor}{2011}]{tay} Taylor, E.N. et al.
2011, MNRAS, 418, 1587
\bibitem[\protect\citeauthoryear{Toomre}{1977}]{toom} Toomre, A. 1977,
{\it Evolution of Galaxies and Stellar Populations, Proceedings
of a Conference and Yale University}, edited by B.M. Tinsley
and R.B. Larson (Yale University Observatory, New Haven, Conn.), 401
\bibitem[\protect\citeauthoryear{Trenti}{2008}]{tren} Trenti, M. \& Stiavelli,
M. 2008, ApJ, 651, 142
\bibitem[\protect\citeauthoryear{Trujillo}{2006}]{truj} Trujillo, I. et al.
2006, ApJ, 650, 18
\bibitem[\protect\citeauthoryear{Valiante}{2015}]{val} Valiante, E. et al. 2015,
in preparation
\bibitem[\protect\citeauthoryear{Van den Bergh}{1976}]{sid} Van den Berg, S.
1976, ApJ, 206, 883
\bibitem[\protect\citeauthoryear{Van der Wel}{2012}]{van} Van der Wel, A.
et al. 2012, ApJ Suppl. 203, 24.
\bibitem[\protect\citeauthoryear{Van der Wel}{2014}]{van} Van der Wel, A.
et al. 2014, ApJ, 792, L6
\bibitem[\protect\citeauthoryear{Windhorst}{2011}]{win} Windhorst, R.A. et al.
2011, ApJ, 193, 27
\bibitem[\protect\citeauthoryear{Wuyts}{2011}]{wuy} Wuyts, S. et al. 2011,
ApJ, 742, 20




\end{thebibliography}
\end{document}